# Turbulence Dynamics based on Lagrange Mechanics and Geometrical Field Theory of Deformation

XIAO Jian-hua

Henan Polytechnic University, Jiaozuo City, China, 454000

**CONTENTS**

**Abstract:** The turbulence field is stacked on the laminar flow. In this research, the laminar flow is described as a macro deformation which forms an instant curvature space. On such a curvature space, the turbulence is viewed as a micro deformation. So, the fluid flow is described by the geometrical field theory of finite deformation. Based on the Lagrange mechanics and the deformation energy concept, using the Least Action Principle, the Euler-Lagrange motion equations are obtained. According to A E Green formulation, the stress concept is introduced by deformation tensor. The fluid motion is described by the multiplication of a macro deformation tensor and a micro deformation tensor. By this way, the geometrical field of fluid motion is well constructed. Then, the spatial derivative of deformation energy is expressed by the gradient of deformation tensors. By this way, the deformation energy related items in the Euler-Lagrange motion equations are expressed by the stress tensor and deformation tensor. The obtained Euler-Lagrange motion equations, then, are decomposed into average deformation equations and turbulence equations. For several special cases, the new results are compared with the conventional Navier-Stokes equation with Reynolds stress modification. By theses comparisons, the intrinsic meaning of Reynolds stress concept is cleared with strict formulation. The comparisons also show that the Bernoulli Equation is a natural precondition for the conventional Navier-Stokes equation. These results explain why there are so many possible understanding about the Reynolds stress. Furthermore, the research confirms that the Navier-Stokes equation with Reynolds stress modification does work. On this sense, the new formulation shows that classical results are first order approximation about the average deformation. As the classical fluid motion equations are the first order approximation, the turbulence wave motion equations are obtained under the condition that the first order approximation is met by a known macro deformation tensor. Generally, the turbulence wave is an inward-traveling wave. Unlike the normal outward-traveling wave related with the average deformation, the inward-traveling wave is the intrinsic feature of turbulence. So, the turbulence is well defined by the equations obtained in this research. For several typical cases, the simplified turbulence wave equations are given out with simple discussion. These results show that, only for viscosity-sensitive turbulence flow, the Reynolds number is reasonable in physical sense. For pressure-sensitive turbulence flow, two kind of Reynolds number are introduced. They fill up the gap for pressure-sensitive turbulence (or high pressure turbulence flow). Furthermore, the vortex flow problems are studied. The vortex flow is classified into two types: pure vortex (volume invariant) and vortex with volume expansion (bubbling vortex). For these very common turbulence phenomena, the turbulence frequency and wave number are completely determined by the macro deformation wave number. These equations are readily to be checked by the available experimental results. To do this work, a special section, comparing with well-known experimental results, is given. The results show that the new formulation as a whole is well supported by the known results. Finally, some problems are discussed at the end of this paper. In one word, turbulence motion equations are obtained and well explained under the theoretic frame of Lagrange mechanics with the help of geometrical field theory of finite deformation, with the help of stress concept derived by deformation energy derivatives. The new theoretic formulation is expected to be valuable for industry applications.

**Key Words:** Reynolds Number, Reynolds stress, turbulence wave equation, Navier-Stokes equation, Bernoulli equation, fluid motion Euler-Lagrange equation, fluid dynamics, Lagrange mechanics, geometrical field

## 1. Introduction

In nature, a vortex street on fluid surface is a very typical fact. When the flow speed is high, bubbles will be produced along with the complicated vortex patterns. If there is a small hole at the bottom of fluid container, the fluid surface will appear a strong vortex on the top position of the hole. These phenomena are mainly controlled by the flow velocity field although the fluid features do have their effects.

Reminding these facts, noting that the Navier-Stokes equation is mainly based on stress concept, the effectiveness of NS equation on explaining such a kind of phenomenon is doubtable. Generally speaking, the mass conservation of the flow plays an important role in the NS equation induction processes. How can we identify a mass element in fluid? As it is clear that micro scale material exchange is common for flow, then the mass conservation in this case are artificial. To get out from this trouble, the mass conservation can be explained as "instant" conservation. Then, how long is this "instant duration"? To answer this question, a "characteristic time scale" must be introduced. Furthermore, as the element size also plays a role in the mass conservation equation, a geometrical size parameter, named as "characteristic length" should be introduced, also. These two parameters are artificial, although it can be argued that, for different fluid material or flow velocity, the characteristic parameters are determined by the flow features [1]. To prove this point, the NS equations are modified by introducing the Reynolds stress. As the Reynolds stress is defined by the correlation of velocity variation, so the correlation character determines the so called characteristic length.

Based on this concept, various Reynolds stress concepts are introduced [2]. The Reynolds number divides the flow into two extreme cases of flow. One is normal flow, another is turbulent flow.

Indeed, by introducing Reynolds stress, much advancement has been achieved. However, the effectiveness of Reynolds stress concept is not universal [3]. The intrinsic reasons of turbulent flow are not answered by the NS equation modified with Reynolds stress satisfactorily [2]. Under such a background, many researches have put their efforts on introducing new equations and proven their results by mathematic modeling of the experimental flow phenomenon [4-8]. These researches reject NS equation with Reynolds stress modification in a roundabout way. If one faces the facts that the main achievements on turbulent flow problem are indeed obtained by the NS equation with Reynolds stress modification, he will still stick on the NS equation by improving the Reynolds stress. This forms another group of researches.

However, the century efforts have only made a limited advancement. Therefore, it is not surprised to be informed by various "news", good or bad.

No matter how energetic arguments are made for NS equations or against NS equations [9], the boundary layer problem (which is found by experimental observation) cannot be solved by the Reynolds stress concept as the boundary has zero velocity. Where, the correlation of velocity lost its objectiveness. In fact, the correlation function usually defines a symmetrical stress tensor. However, the experimental measurements find the asymmetrical stresses. It is because the existence of asymmetrical stress that the boundary layer can be formed. Although asymmetrical stress can be formulated by introducing higher order correlation function of velocity variations, the mechanical and physical reasons are not strong [5].

How to get out from these troubles? Firstly, A S Lodge [10] introduces the body tensor concept to overcome the material objectivity problem. His idea is to define the related physical quantity attached on the material. That is the commoving coordinator system should be used. The material point concept should be attributed to him. However, such an objective has deformation, then the coordinator system is curvature and the curvature is completely determined by the fluid feature and flow field. Mathematically, how to treat such a problem is the key problem. This problem is equivalent with the large deformation problem in solid continuum. C Truesdell [11] treated the large deformation problem by defining the deformation tensor and expressing it as the multiplication of an orthogonal rotation and a stretching tensor. However, this polar decomposition theorem at most only clears the vortex definition problem in fluid dynamics.

Furthermore, the polar decomposition claims that the orthogonal rotation has no contribution to deformation stress as it is equivalent to local rigid rotation [11]. However, for fluid flow, our instinct sense tells us that, the vortex indeed is like the local rigid rotation, while it is this vortex that behaves as very sensitive about the static pressure or dynamic pressure. If the local rigid rotation has no contribution to stress field, how the stress field can control the vortex motion? In fact, mixing the local rigid rotation with the global rigid rotation is the main problem of polar decomposition concept and its related stress (or strain) concept. For dynamic problem, it is clear that the local rigid rotation will produce S-wave. So, the polar decomposition theorem has no value to fluid dynamics problem. On essential sense, in fluid, the linear momentum and the angular momentum are tightly related [12]. Therefore, after finding the commoving coordinators systems is too complicated for fluid flow problem, the main interests return to NS equation with Reynolds stress modification. The research line based on rational mechanics frame is broken down.

To meet the objectiveness of material element under discussion, the traditional way is using the Lagrange coordinator system. Then the Lagrange mechanics (or Hamiltonian mechanics) is used in turbulent flow problem. In such a kind of researches, the Lagrange mechanics is mainly used by mathematical modeling as the general control equations. Therefore, no systematic formulated results are available. This means that the turbulent flow still is a grey box problem.

The main theoretic problems can be summarized as: (1) the mass conservation is doubtable; (2) the linear moment and angular moment are not independent; (3) the Reynolds stress concept is partially justified by experiments, so its correctness should be confirmed partially. However, the Reynolds stress concept should be induced from basic physical laws rather than being introduced by statistic viewpoint; (4) the partial effectiveness of NS equation should be confirmed. This confirmation means that the deformation indeed plays an important role in fluid dynamics problems.

Based on above consideration, this research will formulate the fluid dynamics equations by combining Lagrange mechanics and geometrical field mechanics theory of finite deformation [12]. The basic point is the Lagrange quantity should be defined by the kinetic energy of fluid element and the deformation energy (internal potential energy) of the element. In this sense, the instant Lagrange coordinator system is used. The curvature feature of the commoving dragging coordinator system is determined by the element deformation in local and the global deformation of fluid on average [12]. Therefore, two deformation tensors are introduced. One is the local deformation which determines local velocity gradient referring the global average velocity field. Another is the micro scale deformation tensor which is determined by the subscale velocity variation referring the local average velocity field. In other words, the current velocity field is expressed by the linear transformation of local average velocity field, while the local average velocity field is expressed by the linear transformation of global average velocity field.

To the defined Lagrange quantity, the Least Action principle is used to deduce the Euler-Lagrange equations. Where, the stress concept is introduced by the deformation energy. After establishing the general equations, the stress-deformation relation equations are used, based on classical mechanics consideration. Then, some simple typical flow cases are studied in some details. The related Reynolds numbers are determined by motion equation strictly. These results are compared with the known classical results.

The results of this research show that: the turbulent flow is a semi-deterministic phenomenon in the sense that it is an inward-traveling wave. The cascading phenomenon of vortex flow is deterministic also. The research shows that the Reynolds stress is correct in the sense that the velocity (or exactly, the momentum or kinetic energy) variation indeed plays an equivalent role as the stress in the general motion equations. Furthermore, the research shows that, the general motion equations indeed can be modified as NS equations with suitable stress modification or introducing force item related with velocity field and or deformation.

Special Note: In this research, the geometrical field theory of finite deformation in mechanics will not be explained in details. If the reader wants to known more about it, please download related papers from //arxiv.org, general physics, fluid dynamics; or from //www.paper.edu.cn, mechanics, English version. (The formally published papers do not form a systematic description but only some applications.)

## 2. Stream motion equations based on Lagrange mechanics

Phenomenally, based on traditional fluid dynamics, the deformation energy density $W$ related with the macro deformations is:

$$W = \sigma_j^i \cdot (F_i^j - \delta_i^j) \tag{1}$$

Where, $\sigma_j^i$ is a rank two tensor expressing the stress field; $F_i^j$ is the (point-set transformation) base vector transformation tensor, named as deformation gradient tensor [11]; $\delta_i^j$ is the unit tensor. For this formulation, $(F_i^j - \delta_i^j)$ is defined as the deformation [12]. (For fluid, it corresponds to the instant deformation; For solid, it corresponds to the strain concept in classical elasticity).

However, this definition is not intrinsic. It is a phenomenon expression. So, some further treatment is needed. For Lagrange mechanics, the deformation energy should be used in its general form. By this treatment, the stress will be determined by the energy item in form. This topic is treated by defining the stress as the derivative about strain tensor (A E Green [13]):

$$\sigma_{ij} = \frac{\partial W}{\partial S_{ij}} \tag{2}$$

Where, stress and strain tensors are defined in traditional ways. However, this stress is symmetrical, as the strain is symmetrical. To include the non-symmetrical stress and strain, the definition should be extended. In this research, this definition is extended as:

$$\sigma^i_j = \frac{\partial W}{\partial (F_i^j - \delta_i^j)} \tag{3}$$

Observing this definition, the initial stress for no deformation case must be required as zero. To over come this shortage, some researchers suggest to introduce the initial stress. By this consideration, the stress is defined as:

$$\sigma_i^j = -\frac{\partial U}{\partial F_j^i} \tag{4}$$

Where, $U$ is the internal energy at its current configuration. The negative sign is introduced by the convention of stress is positive for outward surface direction. For no deformation case, $F_j^i = \delta_j^i$, then the initial stress for isotropic fluid material is defined as:

$$\sigma_{0i}^j = -\frac{\partial U}{\partial \delta_j^i} = -p\delta_i^j \tag{5}$$

That means the initial shape is maintained by the pressure. By the stress definition (note that the stress must be defined on the material element), for fluid dynamics, the initial stress is defined as the static pressure at the reference configuration. For solid, the reference configuration is fixed. This initial stress is zero. However, for fluid, the reference configuration is instant. So, the stress-strain relation for fluid will be different from the stress-strain relation for solid. Based on the physical requirement, the static pressure of fluid $p$ will be positive ($p \geq 0$). (Note that the positive energy means the material obtained energy to make its current configuration, so to change the material configuration, some energy must be input to the material. The late is named as the deformation energy. In this research, the zero energy is defined on the reference configuration).

It is well-known that many efforts have been made to establish the deformation mechanics on the bases of thermodynamics [14-15]. So, the above research shows that, unless the stress is well defined, the deformation energy concept will be very confusing. For fluid with expansion in an isolated environment, the only acceptable physical interpretation is the internal energy density decrease, so the stress will be decreased. However, in classical deformation mechanics, the energy variation is positive as the stress sign direction is positive for expansion deformation. Therefore, it is reasonable to abandon the classical stress sign selection for fluid dynamics. However, in this research, the traditional stress sign selection still will be adopted.

Comparing with the classical stress-strain relation equation, the stress in this research should be written as:

$$\sigma_i^j = -p\delta_i^j + \lambda(F_l^l - 3)\delta_i^j + 2\mu(F_i^j - \delta_i^j) \tag{6}$$

Where, $\lambda, \mu$ is the viscosity of material. They are positive.

Anyway, the internal energy is configuration dependent [14-15]. So, it is a kind of potential. Therefore, the potential field $U$ can be introduced to the continuum mechanics.

Physically and mechanically, although the whole translation and rigid rotation has no contribution to the deformation energy density, the kinetic energy density do coupled with the deformation [16]. Such a coupling relationship can be expressed by the Lagrange quantity written as:

$$L = \frac{1}{2} \rho \cdot V^i V^i - U \tag{7}$$

Where, $\rho$ is mass density. Here, the deformation energy is treated as the potential. The reason is that the deformation energy is configuration dependent. The configuration can be understood as a general position concept.

For an objective material element, the total action is expressed as the integration about time interval and the whole configuration body. The action for a deformable element in commoving coordinator system is expressed as:

$$Action = \int_{t_0}^{t} dt \oint_{\Omega} L d\Omega = \int_{t_0}^{t} dt \oint_{\Omega} (\frac{1}{2} \rho \cdot V^i V^i - U) d\Omega \tag{8}$$

Based on Lagrange mechanics formulation, the variance of Lagrange quantity about commoving coordinators $x^i$ and time $t$ is:

$$\delta L = \rho V^i \cdot \frac{\partial \delta x^i}{\partial t} + (\frac{1}{2} V^l V^l \cdot \frac{\partial \rho}{\partial x^i} - \frac{\partial U}{\partial x^i}) \delta x^i + (\rho V^l \cdot \frac{\partial V^l}{\partial t} + \frac{1}{2} V^l V^l \cdot \frac{\partial \rho}{\partial t} - \frac{\partial U}{\partial t}) \delta t \tag{9}$$

Then, the variance of action gives out the following equation:

$$\delta(Action) = \int_{t_0}^{t} dt \oint_{\Omega} \left( \rho V^i \cdot \frac{\partial \delta x^i}{\partial t} + (\frac{1}{2} V^l V^l \cdot \frac{\partial \rho}{\partial x^i} - \frac{\partial U}{\partial x^i}) \delta x^i + (\rho V^l \cdot \frac{\partial V^l}{\partial t} + \frac{1}{2} V^l V^l \cdot \frac{\partial \rho}{\partial t} - \frac{\partial U}{\partial t}) \delta t \right) d\Omega \tag{10}$$

Taking the partial integration about time for the first item of it, for fixed boundary, the following equation is obtained:

$$\delta(Action) = \int_{t_0}^{t} dt \oint_{\Omega} \left( [-\frac{\partial(\rho V^i)}{\partial t} + \frac{1}{2} V^l V^l \cdot \frac{\partial \rho}{\partial x^i} - \frac{\partial U}{\partial x^i}] \delta x^i + (\rho V^l \cdot \frac{\partial V^l}{\partial t} + \frac{1}{2} V^l V^l \cdot \frac{\partial \rho}{\partial t} - \frac{\partial U}{\partial t}) \delta t \right) d\Omega \tag{11}$$

The least action principle [17] requires that the real configuration and commoving coordinator time variation meets the equation:

$$\delta(Action) = 0 \tag{12}$$

Observing the equation, it is easy to find out that, its **Euler-Lagrange equations** are:

$$-\frac{\partial(\rho V^i)}{\partial t} + \frac{1}{2} V^l V^l \cdot \frac{\partial \rho}{\partial x^i} - \frac{\partial U}{\partial x^i} = 0 \tag{13}$$

$$\rho V^l \cdot \frac{\partial V^l}{\partial t} + \frac{1}{2} V^l V^l \cdot \frac{\partial \rho}{\partial t} - \frac{\partial U}{\partial t} = 0 \tag{14}$$

These equations cannot directly be used to deformation problem as the current position of material element $x^i(t)$ is unknown (for mathematical treatment in other ways please see [18-20]). So, only when the spatial derivative is about the laboratory coordinator system $X^i$, the equation can be treated directly. Now, we turn to transform them into the motion equations for deformation.

Note that, here, the material derivative of linear momentum is rejected. This is a natural result of commoving coordinator selection. In fact, in fluid, the acceleration in a spatial point is calculated by the velocity variation about time variation. The total derivative introducing the velocity gradient is not physically acceptable, as the "colored" material element is the only object to take the measurement and perform the calculation.

Generally, within unit time, referring the initial time position $X^i$ of material element, the unit time late position of material element is:

$$x^i = V^i + X^i \tag{15}$$

In other words, the $X^i$ is the laboratory coordinators of the material element at unit time earlier. The $x^i$ is the instant coordinator of the material element at current. In stress definition, it is treated as the commoving dragging coordinator system.

So, the unit time deformation tensor (instant deformation tensor) is defined as:

$$F_j^i = \frac{\partial x^i}{\partial X^j} = \frac{\partial V^i}{\partial X^j} + \delta_j^i \tag{16}$$

When the velocity is defined in the laboratory coordinator system, the instant deformation tensor is well defined.

Observing the potential which is configuration dependent, it can be defined by the internal energy variation between reference configuration and current configuration. That is to say, the zero potential is defined on the reference configuration. By this consideration, the gradient of potential can be expressed as:

$$\frac{\partial U}{\partial x^i} = \frac{\partial [W(x)-W(X)]}{\partial x^i} = \frac{\partial}{\partial x^i}\left(\frac{\partial W}{\partial F_l^k}(F_l^k - \delta_l^k)\right) \tag{17}$$

Where, $W(x)-W(X) = W(F_j^i)-W(\delta_j^i)$ is the deformation energy density for unit time deformation; $W(x)$ is the material internal energy at instant configuration $x^i(t)$; $W(X)$ is the material internal energy at reference configuration $X^i$ (This configuration is defined unit time earlier, so it is also time dependent). The earlier configuration of unit time is taken as the reference configuration. By this formulation, the reference point of deformation potential is the internal energy of static configuration. For related point-views please see [21-22].

On the other hand, the unit time position variation is:

$$dx^i = \frac{\partial V^i}{\partial X^j}dX^j = (F_j^i - \delta_j^i)dX^j \tag{18}$$

For infinitesimal deformation, the commoving potential gradient is transformed into spatial derivative form as:

$$\frac{\partial U}{\partial x^i} = \frac{\partial [W(x)-W(X)]}{\partial x^i} = \frac{\partial}{\partial X^j}\left(\frac{\partial W}{\partial F_j^i}\right) \tag{19}$$

Then, based on the stress definition Equ.(4), the above equation is expressed as:

$$\frac{\partial U}{\partial x^i} = -\frac{\partial \sigma_i^j}{\partial X^j} \tag{20}$$

Hence, omitting the mass density gradient for infinitesimal instant deformation, the **Euler-Lagrange equations for deformation** expressed by stresses are:

$$-\frac{\partial(\rho V^i)}{\partial t}+\frac{1}{2}V^lV^l\cdot\frac{\partial \rho}{\partial x^i}+\frac{\partial \sigma_i^j}{\partial X^j}=0 \tag{21}$$

$$\rho V^l\cdot\frac{\partial V^l}{\partial t}+\frac{1}{2}V^lV^l\cdot\frac{\partial \rho}{\partial t}-\frac{\partial U}{\partial t}=0 \tag{22}$$

The first equation is identical with the classical deformation equation. In fact it is the Navier-Stokes equation. The second equation is energy rate equation. This equation is missing in classical fluid dynamics. Hence, the Navier-Stokes equation is not complete description about fluid dynamics. This problem is caused by the static deformation mechanics, where, the energy rate equation is met automatically.

By the second equation, the deformation energy density rate is time-dependent. This dependence is mainly determined by the velocity variation. It can be inferred that the main problem of classical fluid dynamics is no definite equation for relating the deformation evolution and the velocity evolution. The energy rate equation shows that the kinetic energy can be changed into deformation energy (this is well understood in classical fluid dynamics). However, it also shows that the deformation energy can be changed into kinetic energy (this is well explained by the cracking, Earthquake, and sudden instability of fluid). The late point canon be explained by the classical fluid dynamics equations.

Theoretically, the Equ.(22) are based on the least action principle. It should be more general than the traditional form. It attributes the motion be caused by the deformation, while the stress tensor is waiting to be defined based on physical energy consideration. That is to say, the stress tensor becomes "parameters" determined by deformation tensor.

Furthermore, for time-invariant velocity field, the Euler-Lagrange equations for deformation are simplified as the static deformation equations as:

$$\frac{1}{2}V^l V^l \cdot \frac{\partial \rho}{\partial X^i} + \frac{\partial \sigma_i^j}{\partial X^i} = 0 \tag{23}$$

$$\frac{1}{2}V^l V^l \cdot \frac{\partial \rho}{\partial t} - \frac{\partial U}{\partial t} = 0 \tag{24}$$

This equation shows that the mass density variation is related with the deformation. Based on statistical physics the mass density variation can be expressed by the absolute temperature variation as:

$$\delta\rho = \alpha \cdot \delta T \tag{25}$$

Where, $\alpha$ is a material constant; $T$ is the absolute temperature. The thermo-elastic deformation equations are:

$$\frac{\alpha}{2}V^l V^l \cdot \frac{\partial T}{\partial X^i} + \frac{\partial \sigma_i^j}{\partial X^i} = 0 \tag{26}$$

$$\frac{\alpha}{2}V^l V^l \cdot \frac{\partial T}{\partial t} - \frac{\partial U}{\partial t} = 0 \tag{27}$$

Observing these equations, one finds that the deformation is tightly related with temperature through the velocity field. This temperature dependency of deformation is treated by introducing the temperature stress in classical fluid dynamics. Based on this research, the temperature sensitive stress $\tilde{\sigma}_i^j$ should be defined as:

$$\tilde{\sigma}_i^j(b) = \tilde{\sigma}_i^j(a) - \int_a^b \frac{\alpha}{2}V^l V^l \frac{\partial T}{\partial X^i} dX^j \tag{28}$$

There are many researches about the definition of temperature sensitive stress in statistic physical sense.

Theoretically, once the velocity field $V^i(X)$ is determined, the stream motion is determined. However, the turbulent flow has no contribution to the average velocity. So, the stream motion equations cannot be directly be used to turbulent flow. Although huge efforts have been made to solve turbulent flow problem by means of the Navier-Stokes equation, little intrinsic theoretical advancement has been obtained. Based on observations, the key problem is that the turbulent motion is happened on the stream flow background. Therefore, a set of motion equations that takes the stream motion as reference motion should be found out. How to do this is the main topic bellow.

For the following discussion, the essential condition is that the stream motions are given as known quantity. Hence, the stream motion velocity field is taken as the coordinator of commoving dragging coordinator system.

To do so, the geometrical description problem will be discussed firstly.

## 3. Geometrical description of local material motion

Generally speaking, based on experience working on fluid dynamics problems, the above Euler-Lagrange equations are not complete description of fluid motion. The velocity components which have no contribution to average velocity field do exist. So, local velocity variation is stacked on the stream motion [23]. How to treat this problem is the main target for this research. Firstly, referring to the stream motion, the geometrical description of local material motion will be discussed.

For fluids, the local material motion can be described by the global average velocity field $V^i$ and the local variation velocity field $v^i$, measured in standard rectangular laboratory coordinator system. Both of them are the functions of time $t$ and position $(X^1, X^2, X^3)$. For any material point of fluid, the velocity field can be expressed as:

$$\widetilde{V}^i = V^i + v^i \tag{29}$$

Depending on the characteristic length selection, the global average velocity field $V^i$ represents the fluid configuration variation. Based on classical fluid dynamics theory, the deformation of fluid configuration is described by the strain rate defined as:

$$\varepsilon_{ij} = \frac{1}{2}(\frac{\partial V^i}{\partial X^j} + \frac{\partial V^j}{\partial X^i}) \tag{30}$$

It represents the distance variation of two material elements within unit time. However, such a kind of material element must be understood as an average scale element which is bigger than or equal to the characteristic length (will be referred as macro scale, hereafter).

Within the unit time, the instant deformation tensor can be defined as:

$$F^i_j = \frac{\partial V^i}{\partial X^j} + \delta^i_j \tag{31}$$

Where, $\delta^i_j$ is Kronecher sign. It represents the relative deformation of material element over characteristic length.

Viewing the deformation of material elements as a point sets transformation, a local Lagrange coordinator system can be established, taking the material element center as the reference point. This local coordinator system is attached with the special material element, so it is a co-moving dragging coordinator system [12] and it is represented as $(x^1, x^2, x^3)$, hereafter. In general case, the current configuration is curvature. Referring this current configuration, within the average scale (will be referred as micro scale, hereafter), a local instant deformation tensor can be defined as:

$$G^i_j = v^i\big|_j + \delta^i_j \tag{32}$$

Where, $v^i\big|_j$ represents the co-variant derivative of local velocity variation field about $X^j$ coordinator. In other words, the local dragging coordinator system is determined by the macro local deformation of fluid [12].

Therefore, for a fluid material element, its motion is described by the macro scale deformation $F^i_j(X,t)$ and the micro scale deformation $G^i_j(x,t;X)$. To connecting both of them to the same material element, the co-moving dragging coordinator system should be defined by the current fluid material element configuration. The local current gauge tensor related with the material element under discussion is given by:

$$g_{ij}(X,t) = F_i^l(X,t) \cdot F_j^l(X,t) \tag{33}$$

So, for a material point in micro scale, the deformation is described by the combination of macro scale deformation multiplied by the local deformation as:

$$\widetilde{F}_j^i(x,t;X) = F_l^i(X,t) \cdot G_j^l(x,t;X) \tag{34}$$

This will gives out the current instant gauge tensor as:

$$\widetilde{g}_{ij}(x,t) = \widetilde{F}_i^l(x,t;X) \cdot \widetilde{F}_j^l(x,t;X) = g_{lm}(X,t) G_i^l(x,t;X) G_j^m(x,t;X) \tag{35}$$

However, we should stop here for a while. If one goes too far in this gauge tensor direction, the mathematic problem will be overcome and the physical-mechanics meaning will be demised. For mechanics, the velocity field is the main measurable quantity in fluid motion. So, one should return to velocity field.

After unit time, referring to its initial velocity $V^i(X,t_0) = V_0^i(X)$, the macro scale velocity $V^i(X,t)$ can be expressed as:

$$V^i(X,t) = V_0^i(X) + \frac{\partial V^i}{\partial X^j} \cdot V_0^j(X) = F_j^i(X,t) \cdot V_0^j(X) \tag{36}$$

Note that, after unit time, the material moving distance is $V_0^j$. Similar reasoning is applied to the micro scale velocity field.

Referring to this velocity field, the micro scale deformation is defined. So, for a material point, the velocity after the unit time is described as:

$$\widetilde{V}^i(x,t;X) = G_j^l(x,t;X) \cdot F_l^i(X,t) \cdot V_0^j(X) \tag{37}$$

At time $t_0$, taking the laboratory coordinator system as the co-moving dragging coordinator system, within unit time, their difference can be omitted. Hence, Equ.(37) can be simplified as:

$$\widetilde{V}^i(X,t) = G_j^l(X,t) \cdot F_l^i(X,t) \cdot V_0^j(X) \tag{38}$$

Physically speaking, the velocity field is expressed as: referring to the velocity field at time $t_0$, the current velocity field (unit time later) is determined by the multiplication of macro deformation of fluid material element and the micro deformation within the element. As the reference time $t_0$ is an arbitral choice, the instant velocity field after unit time is expressed referring to the velocity field unit time before.

Based on above geometrical method, the local material motion is described by instant local velocity transformation $\widetilde{F}_j^i(X,t)$, which is defined as:

$$\widetilde{F}_j^i(X,t) = G_j^l(X,t) \cdot F_l^i(X,t) \tag{39}$$

Where, $F_j^i = \frac{\partial V^i}{\partial X^j} + \delta_j^i$ is calculated in laboratory coordinator system directly. However, the $G_j^i = v^i\big|_j + \delta_j^i$ is calculated referring to the current element configuration. Therefore, the different selection of characteristic length will produce different macro and micro transformation, while the instant local velocity transformation $\widetilde{F}_j^i(X,t)$ should be unique. This simply is a physical requirement.

In traditional fluid dynamics, the stress related with macro deformations is the average stress field defined as:

$$\sigma^i_j = -p\delta^i_j + \lambda \cdot (\frac{\partial V^l}{\partial X^l}) \cdot \delta^i_j + 2\mu \cdot \frac{\partial V^i}{\partial X^j} = -p\delta^i_j + \lambda \cdot (F^l_l - 3)\delta^i_j + 2\mu \cdot (F^i_j - \delta^i_j) \quad (40)$$

Where, $p$ is the average static pressure, $\lambda$ and $\mu$ are the viscosity parameters of fluid. All of them are well defined in traditional fluid mechanics. However, in this research, as the stress is introduced by deformation energy gradient about deformation tensor, so their physical implication is energy. Their dimension is energy.

The local relative stress related with the micro deformation is defined as:

$$\tilde{\sigma}^i_j = -\tilde{p}\delta^i_j + \tilde{\lambda} \cdot (v^l\big|_l)\delta^i_j + 2\tilde{\mu} \cdot v^i\big|_j = -\tilde{p}\delta^i_j + \tilde{\lambda} \cdot (G^l_l - 3)\delta^i_j + 2\tilde{\mu} \cdot (G^i_j - \delta^i_j) \quad (41)$$

Where, the $\tilde{p}$ is the local static pressure, $\tilde{\lambda}$ and $\tilde{\mu}$ are the local viscosity parameters of fluid. Unlike the average parameters, they are local parameters and, usually, depend on the local temperature. In many researches, they are named as the dynamic parameters in many researches. For the case $G^i_j = \delta^i_j$, the physical requirement is $\tilde{p} = p$, $\tilde{\lambda} = \lambda$, $\tilde{\mu} = \mu$. By this sense, the ($p, \lambda, \mu$) are global parameters while the ($\tilde{p}, \tilde{\lambda}, \tilde{\mu}$) are local parameters.

Note that the average stress is defined referring to the laboratory coordinator system (initial configuration), while the local relative stress is defined referring to the deformed current configuration, hence, they are not addible directly. So, in general sense, the measured stress variation in laboratory coordinator system does not represent the true local relative stress, as the referring configurations are different.

## 4. Turbulent motion equations based on Lagrange mechanics

Based on above research, the turbulent flow is described by the local micro instant deformation. The local relative deformation energy density related with the micro deformations is:

$$\tilde{U} = \tilde{\sigma}^i_j \cdot (G^j_i - \delta^j_i) \quad (42)$$

Physically and mechanically, although the whole translation and rigid rotation has no contribution to the deformation energy density, the kinetic energy density do coupled with the deformation. Such a coupling relationship can be expressed by the Lagrange quantity written as:

$$L = \frac{1}{2}\rho \cdot \tilde{V}^i \tilde{V}^i - (U + \tilde{U}) \quad (43)$$

Where, $\rho$ is mass density.

According to the definition of stress in deformation mechanics of this research (Equ.(4)), the stress of micro deformation is given by the following definition:

$$\tilde{\sigma}^i_j = -\frac{\partial \tilde{U}}{\partial G^j_i} \quad (44)$$

This stress is referring to the stream line-formed configuration. For fluid macro deformation, the local curvature cannot be omitted. Then, the following results will be obtained.

Based on the geometrical formulation in the third section of this paper, the motion quantities are the two deformation tensors. The micro deformation can be viewed as the variation on the macro deformation. Based on this point, the variance of Lagrange quantity about commoving coordinators is:

$$\delta(Action) = \int_{t_0}^{t} dt \oint_{\Omega} \left( \left(-\frac{\partial(\rho \tilde{V}^i)}{\partial t} + \frac{1}{2}\tilde{V}^l \tilde{V}^l \cdot \frac{\partial \rho}{\partial x^i} - \frac{\partial U}{\partial x^i} - \frac{\partial \tilde{U}}{\partial x^i} \right) \delta x^i + \left(\rho \tilde{V}^l \cdot \frac{\partial \tilde{V}^l}{\partial t} + \frac{1}{2}\tilde{V}^l \tilde{V}^l \cdot \frac{\partial \rho}{\partial t} - \frac{\partial U}{\partial t} - \frac{\partial \tilde{U}}{\partial t} \right) \delta t \right) d\Omega +$$

$$\int_{t_0}^{t} dt \oint_{\Omega} \left( \frac{\rho}{2} \frac{\partial(\tilde{V}^l \tilde{V}^l)}{\partial x^i} - \frac{\partial U}{\partial F_m^l} \frac{\partial F_m^l}{\partial x^i} - \frac{\partial \tilde{U}}{\partial G_m^l} \frac{\partial G_m^l}{\partial x^i} \right) \delta x^i d\Omega \tag{45}$$

The second item is only contributed by the deformation tensors.

So, according to the Lagrange mechanics, its **Euler-Lagrange equations** are:

$$-\frac{\partial(\rho \tilde{V}^i)}{\partial t} + \frac{1}{2}\tilde{V}^l \tilde{V}^l \cdot \frac{\partial \rho}{\partial x^i} - \frac{\partial U}{\partial x^i} - \frac{\partial \tilde{U}}{\partial x^i} + \frac{\rho}{2} \frac{\partial(\tilde{V}^l \tilde{V}^l)}{\partial x^i} - \frac{\partial U}{\partial F_m^l}\frac{\partial F_m^l}{dx^i} - \frac{\partial \tilde{U}}{\partial G_m^l}\frac{\partial G_m^l}{dx^i} = 0 \tag{46}$$

$$\rho \tilde{V}^l \cdot \frac{\partial \tilde{V}^l}{\partial t} + \frac{1}{2}\tilde{V}^l \tilde{V}^l \cdot \frac{\partial \rho}{\partial t} - \frac{\partial U}{\partial t} - \frac{\partial \tilde{U}}{\partial t} = 0 \tag{47}$$

They show that the turbulent stress is coupled with the stream stress through the instant velocity. Referring to spatial parameter and current velocity, based on the physical meaning of this item, noting that $V^i = F_j^i V_0^j$, $\tilde{V}^i = G_j^l F_l^i V_0^j = \tilde{F}_j^i V_0^j$, (Equ.(36)&(38)), the related velocity items are:

$$\tilde{V}^l \tilde{V}^l = \tilde{F}_n^l \tilde{F}_m^l V_0^n V_0^m \tag{48}$$

Following the similar procedure, the **Euler-Lagrange equations** expressed by stresses are:

$$-\frac{\partial(\rho \tilde{V}^i)}{\partial t} + \frac{1}{2}\tilde{V}^l \tilde{V}^l \cdot \frac{\partial \rho}{\partial X^i} + \frac{\partial \sigma_i^j}{\partial X^j} + \frac{\partial \tilde{\sigma}_i^j}{\partial X^j} + \frac{\rho}{2}V_0^m V_0^n \frac{\partial(\tilde{F}_m^l \tilde{F}_n^l)}{\partial X^i} + \sigma_l^m \frac{\partial F_m^l}{\partial X^i} + \tilde{\sigma}_l^m \frac{\partial G_m^l}{\partial X^i} = 0 \tag{49}$$

$$\rho \tilde{V}^l \cdot \frac{\partial \tilde{V}^l}{\partial t} + \frac{1}{2}\tilde{V}^l \tilde{V}^l \cdot \frac{\partial \rho}{\partial t} - \frac{\partial U}{\partial t} - \frac{\partial \tilde{U}}{\partial t} = 0 \tag{50}$$

For infinitesimal deformations, they returns to the traditional equations (Equ.(23-24)). To further simplify the equations, some typical flow cases will be discussed bellow.

### 4.1 No turbulent flow

If there is no micro scale deformation ($G_j^i = \delta_j^i$), the average flow meets equation:

$$-V_0^l \frac{\partial(\rho F_l^i)}{\partial t} + \frac{1}{2}V_0^n V_0^m F_n^l F_m^l \cdot \frac{\partial \rho}{\partial X^i} + \frac{\partial \sigma_i^j}{\partial X^j} + \frac{\rho}{2}V_0^n V_0^m \frac{\partial(F_n^l F_m^l)}{\partial X^i} + \sigma_l^m \frac{\partial F_m^l}{\partial X^i} = 0 \tag{51}$$

$$\rho V_0^n V_0^m F_n^l \cdot \frac{\partial F_m^l}{\partial t} + \frac{1}{2}V_0^n V_0^m F_n^l F_m^l \cdot \frac{\partial \rho}{\partial t} - \frac{\partial U}{\partial t} = 0 \tag{52}$$

Where, $V_0^j$ is the velocity field of a global configuration which is independent of time-space and is taken as the global reference configuration. Making such a choice is to expose the physical meaning of deformation.

If the material is incompressible, the mass density will be constant (In commoving dragging coordinator system, the objectivity of material requires the mass density be constant. This condition means that there is no chemical reaction, also). In this case, it gives out the motion equation:

$$-\rho V_0^l \frac{\partial F_l^i}{\partial t} + \frac{\partial \sigma_i^j}{\partial X^j} + \frac{\rho}{2}V_0^n V_0^m \frac{\partial(F_n^l F_m^l)}{\partial X^i} + \sigma_l^m \frac{\partial F_m^l}{\partial X^i} = 0 \tag{53}$$

$$\rho V_0^n V_0^m F_n^l \cdot \frac{\partial F_m^l}{\partial t} - \frac{\partial U}{\partial t} = 0 \tag{54}$$

In fact, above equations are the motion equations for large scale average deformation flow.

Omitting the deformation tensor spatial contribution from equations Equ.(51-52), the traditional infinitesimal motion equations are:

$$-\frac{\partial(\rho V^i)}{\partial t} + \frac{\partial \sigma_i^j}{\partial X^j} = 0 \tag{55}$$

$$\frac{1}{2}\frac{\partial(\rho V^l V^l)}{\partial t} - \frac{\partial U}{\partial t} = 0 \tag{56}$$

It says that the traditional motion equation Equ.(55) is correct only for constant mass and deformation energy system. It is clear that even in this case, the traditional motion equations are not exact. Missing the kinetic energy and deformation energy relation equation is the intrinsic problem of Navier-Stokes type motion equations.

The equation Equ.(53) can be further rewritten as:

$$-\rho V_0^l \frac{\partial F_l^i}{\partial t} + \frac{\partial \sigma_i^j}{\partial X^j} + (\rho V_0^n V_0^m F_n^l + \sigma_l^m)\frac{\partial F_m^l}{\partial X^i} = 0 \tag{57}$$

It interprets the average stress correction item as the cross correlation determined by the initial average momentum, current deformation, and current stress. This fact is verified by many researches. That is to say, if the Lagrange mechanics is used to fluid dynamics, the Reynolds stress interpretation will be very natural. In fact, this equation is the NS equation with Reynolds stress modification (See Appendix A).

Based on this result, it can be concluded that the applicability condition of traditional motion equation is that the stress meets the following condition:

$$(\rho V_0^n V_0^m F_n^l + \sigma_l^m)\frac{\partial F_m^l}{\partial X^i} \approx 0 \tag{58}$$

It means that: (1) infinitesimal deformation; or (2) the stress is determined purely by the velocity field, deformation tensor, and mass density. The second condition is:

$$\sigma_l^m \approx -\rho V_0^n V_0^m F_n^l \tag{59}$$

In essential sense, it is similar with the Bernoulli Equations [24], if the reference energy is well selected. Note that here the commoving coordinator system is selected. This point will be cleared more, later.

If one still uses the viscosity parameters $\lambda, \mu$, and deformation tensor to express the stress, one must conclude that the viscosity parameters $\lambda, \mu$ are velocity dependent. This just is what happed in last several decades.

In fact, in airplane flying problems, the stress indeed is defined similar with equation Equ.(59). However, this topic is a little too far for present discussion. So, we will return to the main line of this research.

**4.2 Constant average flow.**

In fact, for turbulent flow, the average flow is near constant ($\frac{\partial F_j^i}{\partial t} = 0$), so the Euler-Lagrange equations will lead to the turbulent flow motion equations expressed as:

$$-V_0^m F_m^n \frac{\partial(\rho G_n^i)}{\partial t} + \frac{1}{2}V_0^m V_0^n \tilde{g}_{mn}\frac{\partial \rho}{\partial X^i} + \frac{\partial \sigma_i^j}{\partial X^j} + \frac{\partial \tilde{\sigma}_i^j}{\partial X^j} + \frac{\rho}{2}V_0^n V_0^m \frac{\partial \tilde{g}_{nm}}{\partial X^i} + \sigma_l^m \frac{\partial F_m^l}{\partial X^i} + \tilde{\sigma}_l^m \frac{\partial G_m^l}{\partial X^i} = 0 \tag{60}$$

$$\rho V_0^m V_0^n \frac{\partial \tilde{g}_{mn}}{\partial t} + \frac{1}{2} V_0^m V_0^n \tilde{g}_{mn} \cdot \frac{\partial \rho}{\partial t} - \frac{\partial U}{\partial t} - \frac{\partial \tilde{U}}{\partial t} = 0 \tag{61}$$

Where, the current gauge tensor $\tilde{g}_{ij}$ is defined as:

$$\tilde{g}_{ij} = \tilde{F}_i^l \tilde{F}_j^l = (G_i^m F_m^l)(G_j^n F_n^l) = (G_i^m G_j^n)(F_m^l F_n^l) = G_i^m G_j^n g_{mn} \tag{62}$$

For constant average flow, based on experimental results, the turbulent flow has no significant contribution to the average gauge tensor. So, on average sense, the above equation Equ.(62) can be approximated as:

$$\tilde{g}_{ij} \approx g_{mn} = F_i^l F_j^l \tag{63}$$

Based on this approximation, the turbulent flow motion equations are:

$$-V_0^m F_m^n \frac{\partial (\rho G_n^i)}{\partial t} + \frac{1}{2} V_0^m V_0^n \frac{\partial (\rho g_{mn})}{\partial t} + \frac{\partial \sigma_i^j}{\partial X^j} + \frac{\partial \tilde{\sigma}_i^j}{\partial X^j} + \sigma_l^m \frac{\partial F_m^l}{\partial X^i} + \tilde{\sigma}_l^m \frac{\partial G_m^l}{\partial X^i} = 0 \tag{64}$$

$$\frac{1}{2} V_0^m V_0^n g_{mn} \cdot \frac{\partial \rho}{\partial t} - \frac{\partial U}{\partial t} - \frac{\partial \tilde{U}}{\partial t} = 0 \tag{65}$$

If the mass density is constant, the equations Equ.(64-65) can be simplified as:

$$\rho V_0^m F_m^n \frac{\partial G_n^i}{\partial t} - \frac{\rho}{2} V_0^m V_0^n \frac{\partial g_{mn}}{\partial X^i} - \frac{\partial \sigma_i^j}{\partial X^j} - \frac{\partial \tilde{\sigma}_i^j}{\partial X^j} - \sigma_l^m \frac{\partial F_m^l}{\partial X^i} - \tilde{\sigma}_l^m \frac{\partial G_m^l}{\partial X^i} = 0 \tag{66}$$

$$\frac{\partial U}{\partial t} + \frac{\partial \tilde{U}}{\partial t} = 0 \tag{67}$$

In general sense, for a constant deformation, its deformation energy should be constant. However, the equation Equ.(67) says that this is not true for turbulent flow, unless the turbulent flow is constant. However, the constant turbulent flow contradicts with the observed turbulent production processes. The observed results show that the macro deformation energy and turbulence deformation energy is interchangeable.

In fact, the equation Equ.(67) shows the energy can be circulating between the average flow and turbulent flow by the total energy conservation. That is:

$$U + \tilde{U} = U_0 (Const) \tag{68}$$

The total energy conservation for fully developed turbulent flow is a fact verified by many experiments. The deformation energy conservation shows that the turbulent flow will control the macro deformation or the macro viscosity. The late is referred as dynamic viscosity or non-linear viscosity.

In some special cases (especially in short time duration after the turbulent flow is fully developed), the constant turbulent flow does exist. So, the constant turbulent flow equations are:

$$\frac{\rho}{2} V_0^m V_0^n \frac{\partial g_{mn}}{\partial X^i} + \sigma_l^m \frac{\partial F_m^l}{\partial X^i} + \tilde{\sigma}_l^m \frac{\partial G_m^l}{\partial X^i} = 0 \tag{69}$$

It is the intrinsic reason for that, based on traditional mechanics, for a fully developed turbulent flow, the stresses are balanced.

$$\rho V_0^m F_m^n \frac{\partial G_n^i}{\partial t} - \frac{\partial \sigma_i^j}{\partial X^j} - \frac{\partial \tilde{\sigma}_i^j}{\partial X^j} = 0 \tag{70}$$

In an approximation sense, the equation Equ.(69) allows such a kind of solution:

$$\frac{\rho}{2} V_0^m V_0^n g_{mn} + \sigma_l^m F_m^l + \tilde{\sigma}_l^m G_m^l = Const \tag{71}$$

It shows that for fully developed turbulent flow, the turbulent deformation and the average stream deformation have a deterministic relationship. This fact causes many confusing in previous researches as the Reynolds stress is introduced by non-deterministic methods. This phenomenon is referred as pattern. However, based on this research, it only is one special solution of Equ(69).

For turbulent production problems, based on physical consideration that the initial turbulence flow stress is very small, the turbulent production equations should be:

$$\rho V_0^m F_m^n \frac{\partial G_n^i}{\partial t} - \frac{\rho}{2} V_0^m V_0^n \frac{\partial g_{mn}}{\partial X^i} - \frac{\partial \sigma_i^j}{\partial X^j} - \sigma_l^m \frac{\partial F_m^l}{\partial X^i} = 0 \tag{72}$$

$$\frac{1}{2} V_0^m V_0^n g_{mn} \cdot \frac{\partial \rho}{\partial t} - \frac{\partial U}{\partial t} - \frac{\partial \tilde{U}}{\partial t} = 0 \tag{73}$$

Where, the macro deformation is taken as known quantity. They show that there are three intrinsic causes for turbulent flow production: (1) average steam deformation (usually appears in high curvature streamline); (2) macro deformation energy or viscosity parameters variation (usually appears in multi-phase processes); and (3) mass density variation (usually appears in multi-composition flow processes).

If the equation Equ.(69) is met, the equation Equ.(49) can be simplified as:

$$-\frac{\partial(\rho \tilde{V}^i)}{\partial t} + \frac{1}{2} \tilde{V}^l \tilde{V}^l \cdot \frac{\partial \rho}{\partial X^i} + \frac{\partial \sigma_i^j}{\partial X^j} + \frac{\partial \tilde{\sigma}_i^j}{\partial X^j} = 0 \tag{74}$$

Observing the results and comparing with the results of no turbulence flow, it seems that the turbulent flow problem still can be solved by the traditional flow equations with Reynolds stress modification. However, the fact of complexity of turbulent flow tells us that the correct Reynolds stress definition should be:

$$\frac{\partial \sigma_{Re\,i}^j}{\partial X^j} = \frac{\partial \sigma_i^j}{\partial X^j} + \sigma_l^m \frac{\partial F_m^l}{\partial X^i} + \tilde{\sigma}_l^m \frac{\partial G_m^l}{\partial X^i} + \frac{1}{2} V_0^m V_0^n g_{mn} \frac{\partial \rho}{\partial X^i} \tag{75}$$

Therefore, Reynolds stress correction concept is correct although the correction methods are not exact. This interprets why so many efforts are put on the Reynolds stress definition or calculation problems.

**4.3 Incompressible fluid**

In mechanics research, a wide degree of fluid material can be approximated by the incompressible material. The incompressible condition is expressed as:

$$\tilde{g}_{ij} = \delta_{ij}, \quad g_{ij} = \delta_{ij} \tag{76}$$

Based on the previous geometrical formulation, this condition means that:

$$F_i^l F_j^l = R_i^l R_j^l = \delta_{ij}, \quad G_i^l G_j^l = \tilde{R}_i^l \tilde{R}_j^l = \delta_{ij} \tag{77}$$

That is to say both deformations ($F_j^i = R_j^i$, $G_j^i = \tilde{R}_j^i$) are orthogonal local rotation [12,25-26]. In this case, the general motion equations are simplified as:

$$\rho V_0^m \frac{\partial(R_m^n \tilde{R}_n^i)}{\partial t} - \frac{\partial \sigma_i^j}{\partial X^j} - \frac{\partial \tilde{\sigma}_i^j}{\partial X^j} - \sigma_l^m \frac{\partial R_m^l}{\partial X^i} - \tilde{\sigma}_l^m \frac{\partial \tilde{R}_m^l}{\partial X^i} = 0 \tag{78}$$

$$\frac{\partial U}{\partial t} - \frac{\partial \tilde{U}}{\partial t} = 0 \tag{79}$$

So, for incompressible flow problems, the deformation energy conservation is a natural condition.

A balanced macro deformation can be defined as:

$$\rho V_0^m \frac{\partial R_m^i}{\partial t} - \frac{\partial \sigma_i^j}{\partial X^j} = 0 \tag{80}$$

Its solution problems are discussed in [25-26]. It says that the macro local rotation rate determines the stress field. The Reynolds stress should be defined as:

$$\frac{\partial \sigma_{\text{Re}\,i}^j}{\partial X^j} = \sigma_l^m \frac{\partial R_m^l}{\partial X^i} + \tilde{\sigma}_l^m \frac{\partial \tilde{R}_m^l}{\partial X^i} + [\rho V_0^m \frac{\partial (R_m^n \tilde{R}_n^i)}{\partial t} - \rho V_0^m \frac{\partial R_m^i}{\partial t}] \tag{81}$$

The equation Equ.(78), then, is expressed by Reynolds stress with correction item as:

$$\rho V_0^m \frac{\partial (R_m^n \tilde{R}_n^i)}{\partial t} - \frac{\partial (\sigma_i^j + \tilde{\sigma}_i^j)}{\partial X^j} - \frac{\partial \sigma_{\text{Re}\,i}^j}{\partial X^j} = 0 \tag{82}$$

In form, it returns to the traditional fluid motion equation. However, the Reynolds stress correction is defined by equation Equ.(81). Therefore, there are intrinsic differences.

Based on above formulation, a reasonable way to solve turbulence motion equations is: firstly, solve equation Equ.(80) to obtain the local average deformation; secondly, replacing the global average velocity with the obtained local average velocity in first step, solve equation Equ.(81) by suitable consideration about micro deformation; finally, combine the both results to get the complete solutions by equation Equ.(82) to get the micro deformation. Repeating this procedure, high precision results should be available.

### 4.4 Conservative flow with turbulence

Such a kind flow plays an important role in fluid dynamics, as the free flow meets deformation energy conservation condition. Usually, if the reference velocity field $V_0^i$ is not time dependent, that is to say they are only the function of spatial positions, then the flow is steady flow in the sense of characteristic length. For steady flow, in the characteristic length, the total energy density should be conservative. In this case, the turbulent flow motion is defined by the conservation of energy. For idea steady flow without turbulence, as the turbulence is grown up with the time elapsing, so the turbulent motion should meet the energy conservation condition:

$$\rho \tilde{V}^l \cdot \frac{\partial \tilde{V}^l}{\partial t} - \frac{\partial U}{\partial t} - \frac{\partial \tilde{U}}{\partial t} = 0 \tag{83}$$

If there is no turbulence, the energy conservation condition is:

$$\rho V^l \cdot \frac{\partial V^l}{\partial t} - \frac{\partial U}{\partial t} = 0 \tag{84}$$

So, the turbulence energy conservation is expressed as:

$$\tilde{U} = \frac{\rho}{2} V_0^m V_0^n (\tilde{g}_{mn} - g_{mn}) = \frac{\rho}{2} V_0^m V_0^n F_m^k F_n^l (G_k^p G_l^p - \delta_{kl}) \tag{85}$$

It is easy to verify that:

$$(G_k^p G_l^p - \delta_{kl}) \approx \varepsilon_{kp} \varepsilon_{lp} + 2\varepsilon_{kl} \tag{86}$$

Where, the $\varepsilon_{kl} = \frac{1}{2}(G_l^k + G_k^l) - \delta_{kl}$ is the strain definition of turbulence in classical sense.

Therefore, based on traditional deformation energy definition, omitting higher infinitesimal items of asymmetrical components, the turbulence energy is approximated as:

$$\tilde{U} \approx \rho V_0^m V_0^n F_m^k F_n^l \varepsilon_{kl} + \frac{1}{2} \rho V_0^m V_0^n F_m^k F_n^l \varepsilon_{pk} \varepsilon_{pl} \tag{87}$$

It shows that the turbulence energy is completely determined by the macro deformation. The first item is kinetic energy variation caused by deformation. By equation (84), it is the macro deformation energy variation. So, one has:

$$\tilde{\sigma}_k^l = -\rho V_0^m V_0^n F_m^k F_n^l - \rho V_0^m V_0^n F_m^p F_n^l \varepsilon_{pk} \qquad (88)$$

This equation can be viewed as the analytic form of Bernoulli Equation.

Therefore, for conservative flow with turbulence, the turbulence stress is determined by the average flow rather than the turbulence deformation. However, the stress variation which is related with the velocity variation indeed is related with the stress defined by velocity variation correlation. This point verifies the reasonability of Reynolds stress concept. For engineering problems, this result should be very useful. It gives out a way to calculate the Reynolds stress directly by the average velocity field and macro deformation. It simply says that the turbulence elasticity is the kinetic energy dependent. This conclusion is very important for understanding the turbulence phenomenon.

For zero turbulence flow, the motion equation is:

$$-\frac{\partial(\rho V^i)}{\partial t} + \frac{1}{2} V^l V^l \cdot \frac{\partial \rho}{\partial X^i} + \frac{\partial \sigma_i^j}{\partial X^j} + \frac{1}{2} V_0^n V_0^m \frac{\partial(\rho F_n^l F_m^l)}{\partial X^i} + \sigma_l^m \frac{\partial F_m^l}{\partial X^i} = 0 \qquad (89)$$

Introducing the turbulence velocity as:

$$v^i = \tilde{V}^i - V^i = V_0^m F_m^n (G_n^i - \delta_n^i) \qquad (90)$$

Note that, for $|v^i| \ll |V^i|$:

$$\tilde{V}^l \tilde{V}^l - V^l V^l \approx 2 v^l V^l \qquad (91)$$

Taking the macro deformation as known quantity (the solution of zero turbulence motion equation Equ.(89)), the motion equation for conservative flow with turbulence is:

$$-\frac{\partial(\rho v^i)}{\partial t} + v^l V^l \cdot \frac{\partial \rho}{\partial X^i} + \frac{\partial \tilde{\sigma}_i^j}{\partial X^j} + \frac{\partial(\rho v^l V^l)}{\partial X^i} + \tilde{\sigma}_l^m \frac{\partial G_m^l}{\partial X^i} = 0 \qquad (92)$$

Omitting the asymmetrical components, the turbulence motion equation is re-written as:

$$\frac{\partial(\rho v^i)}{\partial t} - \frac{\partial \tilde{\sigma}_i^j}{\partial X^j} = v^l V^l \cdot \frac{\partial \rho}{\partial X^i} + \frac{\partial(\rho v^l V^l)}{\partial X^i} + \tilde{\sigma}_l^m \frac{\partial \varepsilon_{ml}}{\partial X^i} \qquad (93)$$

It shows that the conservative flow with turbulence does not satisfy the classical motion equation.

Note that the turbulence stress can be expressed as:

$$\tilde{\sigma}_k^l = -\rho V^k V^l - \rho V^p V^l \varepsilon_{pl} \qquad (94)$$

It means that, for turbulence region, the dynamic viscosities are purely determined by fluid velocity field. It is this reason that makes the fluid feature definition be arbitral in many researches.

Note that, for micro turbulence, the macro deformation field spatial gradient is viewed as infinitesimal, so, they are omitted. That is to say the macro velocity is viewed as spatial constant within the scale which the turbulence motion is under discussion. Omitting the higher infinitesimal items, the linear turbulence flow motion equation is:

$$\frac{\partial(\rho v^i)}{\partial t} + \left(\rho V^l V^l\right) \frac{\partial \varepsilon_{lj}}{\partial X^j} = v^l V^l \cdot \frac{\partial \rho}{\partial X^i} - \rho V^l \cdot \frac{\partial v^l}{\partial X^i} \qquad (95)$$

Only when the macro deformation is given, the turbulence equation can be solved.

As a simple example, consider one dimensional flow problem ($V^1 = V$, $V^2 = 0$, $V^3 = 0$). If the mass density is constant, the simplest turbulence motion equation is:

$$\frac{\partial v^1}{\partial t} + (V)^2 \left( \frac{\partial^2 v^1}{\partial X^1 \partial X^1} + \frac{\partial^2 v^1}{\partial X^2 \partial X^2} + \frac{\partial^2 v^1}{\partial X^3 \partial X^3} \right) = -V \cdot \frac{\partial v^1}{\partial X^1} \quad (96\text{-}1)$$

$$\frac{\partial v^2}{\partial t} = -V \cdot \frac{\partial v^1}{\partial X^2} \quad (96\text{-}2)$$

$$\frac{\partial v^3}{\partial t} = -V \cdot \frac{\partial v^1}{\partial X^3} \quad (96\text{-}3)$$

Viewing the equation Equ.(96-1), as the turbulence transportation direction is opposite with the flow direction, the turbulence seems being stay at a fixed spatial position. This is the main feature of turbulence phenomenon in every day phenomenon.

Putting Equ.(96-2) and (96-3) into equation Equ.(96-1), one has:

$$\frac{\partial v^1}{\partial t} + (V)^2 \left( \frac{\partial^2 v^1}{\partial X^1 \partial X^1} \right) - V \cdot \left( \frac{\partial^2 v^2}{\partial t \partial X^2} + \frac{\partial^2 v^3}{\partial t \partial X^3} \right) = -V \cdot \frac{\partial v^1}{\partial X^1} \quad (97)$$

It says that the turbulence wave is not stable.

If the turbulence velocity has zero divergence:

$$\frac{\partial v^1}{\partial X^1} + \frac{\partial v^2}{\partial X^2} + \frac{\partial v^3}{\partial X^3} = 0 \quad (98)$$

The equation is simplified as one dimensional wave equation:

$$\frac{\partial v^1}{\partial t} + (V)^2 \left( \frac{\partial^2 v^1}{\partial X^1 \partial X^1} \right) + V \frac{\partial^2 v^1}{\partial t \partial X^1} = -V \cdot \frac{\partial v^1}{\partial X^1} \quad (99)$$

It shows that the time evolution and space evolution features are completely determined by the average flow velocity. Although this equation is highly non-linear, it meets scaling rules.

As the suitable scale is defined by $\frac{\partial V}{\partial X^1} \approx 0$, after simple mathematical operation, the equation Equ.(99) gives out a form solution, which is a turbulence meets the wave motion equation:

$$\frac{\partial v^1}{\partial t} + V \cdot \frac{\partial v^1}{\partial X^1} = \left( \frac{\partial v^1}{\partial t} + V \cdot \frac{\partial v^1}{\partial X^1} \right)\bigg|_{X^1 = X_0} \cdot \exp\left( -\frac{X^1 - X_0}{V} \right) \quad (100\text{-}1)$$

Where, $X_0$ is a reference point. It says that the turbulence wave source item is exponential decaying about spatial distance. For a wave form solution:

$$v^1 = f(X^1 + Vt) \quad (100\text{-}2)$$

The turbulence wave evolution equation is:

$$f'(X^1 + Vt) = f'(X_0 + Vt) \cdot \exp\left( -\frac{X^1 - X_0}{V} \right) \quad (100\text{-}3)$$

Where, $f'(x) = \frac{df(x)}{dx}$. It shows that the turbulence acceleration wave is exponential decaying about distance. So, far away from the turbulence center and/or position, the turbulence tends to be zero. This feature forms the special turbulence distribution pattern. If one views the $f'(x)$ as a wave motion pattern, then it is concluded that the pattern is exponential decaying along with the distance increase. For constant macro velocity, taking Fourier transformation for both sides about time parameter, one will get spectral distribution about spatial position as:

$$F(X^1, \omega) = F(X_0, \omega) \cdot \exp\left(-\frac{X^1 - X_0}{V}\right) \tag{100-4}$$

As the Fourier transformation of spatial correlation function is the spectral density (the square of $|F(X, \omega)|$), so, the spatial correlation variation of turbulence velocity field is exponential decaying about the scale. This leads to the well-known conclusion about turbulence: high correlation within short-distance, low correlation for long distance. This will leads one to define a scale to treat the turbulence problem. If these scaling rules are repeated used to make approximation solutions, the turbulence embedding structure will be established. This procedure has been become very common practices in turbulence researches.

It also shows that the coherent structure of turbulence indeed is the intrinsic feature of turbulence. There are many papers about this topic. Here, there are no needs to further discuss it.

As a first approximation, omitting the higher order derivatives, the linear turbulence wave equation is

$$\frac{\partial v^1}{\partial t} + V \cdot \frac{\partial v^1}{\partial X^1} = 0 \tag{101}$$

In general case, $V = V(X^1, t)$, so it is a non-linear wave equation. For constant $V$, any continuous function $v^1 = f(X^1 + Vt) = const$ is its solution. Hence, the wave traveling direction is inward. The iso-phase line is given by the geometrical condition: $X^1 + Vt = const$.

Unlike the outward-traveling wave, turbulence wave traveling is in-ward direction. It is the reverse process of normal outward-traveling wave. For such a kind of wave, focusing phenomenon is the main feature. In the focus points, the wave amplitude is very high, although the wave amplitude is very low in global sense. The turbulence focus point position is determined by the disturbance near a boundary which is defined by the common phase zone. Therefore, the turbulence in focus may be very high, while the global turbulence is very low. Thus, the focusing feature of turbulence wave equation makes the turbulence in focusing zone seems to be self-grown up suddenly. This is the key point for turbulence research, which is omitted by many researches. The shock wave research pays too much attention on the mathematic reasons, unfortunately, the physical reality is omitted.

Summering up above researches, it is shown that the motion equations obtained in this research can derive out the classical motion equations. Therefore, it can be viewed as supported by classical results. Now, it is the time to turn to the key problem: how to use the new equations to solve the turbulence experiments with higher accuracy than the old research. This will be the main topic for the next section.

## 5. Wave equations of turbulence for known macro flow

In many researches with industrial applications, the macro flow can be viewed as known. The main problem is what kind of turbulence will be produced by such a known macro flow. In mechanics, the turbulence can be viewed as an additional process which is stack on the macro flow. Based on this point, the no-turbulence flow motion equations Equ.(53) and (54) are satisfied. In physical consideration, on average sense, the macro stress should include the turbulence stress correction. So, in approximation sense, the macro deformation solution meets the following equations:

$$-G_l^i \frac{\partial(\rho V^l)}{\partial t} + \frac{1}{2}\tilde{V}^l\tilde{V}^l \cdot \frac{\partial \rho}{\partial X^i} + \frac{\partial \sigma_i^j}{\partial X^j} + \frac{\partial \tilde{\sigma}_i^j}{\partial X^j} + \frac{1}{2}V_0^n V_0^m \frac{\partial(\rho \tilde{F}_n^l \tilde{F}_m^l)}{\partial X^i} = 0 \tag{102}$$

$$\rho \tilde{V}^l \cdot \frac{\partial \tilde{V}^l}{\partial t} + \frac{1}{2} \tilde{V}^l \tilde{V}^l \cdot \frac{\partial \rho}{\partial t} - \frac{\partial U}{\partial t} - \frac{\partial \tilde{U}}{\partial t} \approx 0 \quad (103)$$

Under these conditions, the **Euler-Lagrange equations** Equ.(49) and (50) are used to derive the turbulence wave equations. After simple algebra operation, the **general turbulence wave equations** are obtained as:

$$\rho V^l \frac{\partial G_l^i}{\partial t} - \sigma_l^m \frac{\partial F_m^l}{\partial X^i} - \tilde{\sigma}_l^m \frac{\partial G_m^l}{\partial X^i} = 0 \quad (104)$$

This is the result which will be used late in this research. It attributes the wave motion be caused by the deformation, while the turbulence stress tensor is waiting to be defined based on physical energy consideration. That is to say, the turbulence motion is a non-linear wave phenomenon which takes the stress tensor and macro deformation as its "parameters" and source item.

As the current velocity and stresses are expressed by the deformation tensors and initial velocity, so they supply the general equations to relate the two deformations. To clear its intrinsic meaning, several special cases are discussed bellow.

### 5.1 Turbulence wave equations for simple fluid

For idea simple fluid material, the macro deformation stress tensor is defined as:

$$\sigma_l^m = -p\delta_l^m + \lambda(F_l^l - 3)\delta_l^m + 2\mu(F_l^m - \delta_l^m) \quad (105)$$

Omitting the macro shear viscosity related with fluid elasticity, Equ.(104) is expressed as:

$$\rho V^l \frac{\partial G_l^i}{\partial t} + [(p - \lambda \cdot \Delta) + 2\mu] \cdot \frac{\partial F_l^l}{\partial X^i} - 2\mu F_l^m \frac{\partial F_m^l}{\partial X^i} - \tilde{\sigma}_l^m \frac{\partial G_m^l}{\partial X^i} = 0 \quad (106)$$

Where, $\Delta = F_l^l - 3$ is understood as the macro volume variation. It shows that the micro deformation is determined by the mass density, static pressure, initial velocity, and viscosity parameter $\lambda$. In fact, the viscosity item can be omitted for most high speed or high pressure flow. On phenomenal observation, the micro deformation is produced by the global average velocity and/or mass density variation about space.

Based on Equ.(88), the turbulence stress can be expressed as:

$$\tilde{\sigma}_k^l = -\rho V^k V^l - \rho V^p V^l (G_k^p - \delta_k^p) \quad (108)$$

Omitting the second infinitesimal item, putting it into the equation Equ.(106), one has:

$$\rho V^l \frac{\partial G_l^i}{\partial t} + [(p - \lambda \cdot \Delta) + 2\mu] \cdot \frac{\partial \Delta}{\partial X^i} - 2\mu F_l^m \frac{\partial F_m^l}{\partial X^i} + \rho V^l V^m \frac{\partial G_m^l}{\partial X^i} = 0 \quad (109)$$

This is the **turbulence wave equations for simple fluid.** The striking feature is that: the turbulence wave is a inward-traveling wave, although it is highly non-linear. The effectiveness of wave methods in turbulence research can be supported by this equation. Based on wave theory, the focus point or zone is always possible for random noise appeared in global field.

There are two types of turbulences have very important scientific meaning and industrial application value.

### 5.2 Viscosity sensitive turbulence wave equation

The first type is the viscosity caused turbulence which has been studied extensively and, in fact, is the classical definition of turbulence. Here, following the classical procedure, letting $\Delta = 0$, the turbulence wave equation is simplified as:

$$\rho V^l \frac{\partial G_l^i}{\partial t} - 2\mu F_l^m \frac{\partial F_m^l}{\partial X^i} + \rho V^l V^m \frac{\partial G_m^l}{\partial X^i} = 0 \quad (110)$$

As the force source is the macro deformation with the viscosity as its main parameter, this kind of turbulence should be named as viscosity-sensitive type turbulence. The condition $\Delta = 0$ is named as "incompressible" in classical fluid dynamics. So, it is the main type of turbulence in liquid fluids which has been well-studied until now.

**5.3 Pressure sensitive turbulence wave equation**

The second type is the pressure caused turbulence which has not been studied extensively, although there are many experimental reports explained the observed results. For many turbulent researches and experiments, the conditions $\lambda \Delta = \lambda (F_l^l - 3) \ll p$ are approximated by the fluid flow feature. For this case, it can be further simplified as:

$$\rho V^l \frac{\partial G_l^i}{\partial t} + (p + 2\mu) \cdot \frac{\partial F_l^l}{\partial X^i} - 2\mu F_l^m \frac{\partial F_m^l}{\partial X^i} + \rho V^l V^m \frac{\partial G_m^l}{\partial X^i} = 0 \tag{111}$$

For macro deformation under very high pressure sense, $2\mu \ll p$, the simplest turbulence wave equation is obtained as:

$$\rho V^l \frac{\partial G_l^i}{\partial t} + p \cdot \frac{\partial F_l^l}{\partial X^i} + \rho V^l V^m \frac{\partial G_m^l}{\partial X^i} = 0 \tag{112}$$

This is the pressure sensitive turbulence wave equation. It has no relation with viscosity. The only three parameters are: mass density, pressure, and macro velocity.

As the force source is the macro deformation with the pressure as its main parameter, this kind of turbulence should be named as pressure-sensitive type turbulence. The condition $2\mu \ll p$ is named as extremely low viscosity fluid in classical fluid dynamics. So, it is the main type of turbulence in high pressure aerodynamics which has been well-studied until now.

Even for this simplest case, there are several special cases which have experimental observation importance which will be the main topic for the next two sections.

## 6. Wave equations of viscosity-sensitive turbulence

In many researches, the turbulence is viewed as a kid of wave with random feature but not completely random. The viscosity-sensitive turbulence is produced under the condition $\Delta = F_l^l - 3 = 0$. Its wave equation is Equ.(111). Based on classical mechanics, the condition ($\Delta = F_l^l - 3 = 0$) means that the macro deformation is purely shear deformation. Very low velocity and very high velocity cases will be discussed. The experiments will be examined when it is appropriate.

**6.1 Very low speed flow**

Very low speed flow means that:

$$\left| 2\mu F_l^m \frac{\partial F_m^l}{\partial X^i} \right| \gg \left| \rho V^l V^m \frac{\partial G_m^l}{\partial X^i} \right| \tag{113}$$

In this case, the wave equation Equ.(111) is simplified as:

$$\rho V^l \frac{\partial G_l^i}{\partial t} - 2\mu F_l^m \frac{\partial F_m^l}{\partial X^i} = 0 \tag{114}$$

Its form solution is:

$$\rho V^l G_l^i = 2\mu \int_0^t F_l^m \cdot \frac{\partial F_m^l}{\partial X^i} dt + \rho V^l G_l^i \Big|_{t=0} \tag{115}$$

The instant velocity is completely deterministic. They are:

$$\tilde{V}^i = \frac{2\mu}{\rho} \int_0^t F_l^m \cdot \frac{\partial F_m^l}{\partial X^i} dt + \tilde{V}^i \Big|_{t=0} \tag{116}$$

It shows that the flow velocity variation is controlled by the viscosity, mass density, and macro deformation gradient. This solution is very valuable for industrial application. It shows that, by controlling the stream line or tube shape, the turbulence amplitude can be controlled. In fact, this topic has been a hot topic in fluid researches.

**6.2 Very high speed flow**

Very high speed flow means that:

$$\left| 2\mu F_l^m \frac{\partial F_m^l}{\partial X^i} \right| << \left| \rho V^l V^m \frac{\partial G_m^l}{\partial X^i} \right| \tag{117}$$

In this case, the wave equation Equ.(111) is simplified as:

$$V^l \frac{\partial G_l^i}{\partial t} + V^l V^m \frac{\partial G_m^l}{\partial X^i} = 0 \tag{118}$$

This kind of inward-traveling wave has anisotropy transportation speed. It has three independent components equations. Therefore, generally speaking, the micro deformation cannot be completely determined. However, as usual practice in industry, expressing the macro deformation tensor by its velocity form and/or introducing the symmetry condition, the Equ.(118) becomes a solvable wave equation.

For one dimensional case, the focusing wave velocity simply is the flow velocity.

$$\frac{\partial G_1^1}{\partial t} = -V^1 \frac{\partial G_1^1}{\partial X^1} \tag{119}$$

On phenomenon, it seems that the turbulence is fixed at the same spatial position. Or in exact words, the turbulence is a standing wave.

For general case, there are three independent wave components equations. Each has a different transportation speed. For symmetrical case, one has:

$$V^l \frac{\partial \varepsilon_{il}}{\partial t} + V^l V^m \frac{\partial \varepsilon_{lm}}{\partial X^i} = 0, \quad \varepsilon_{il} = \frac{1}{2}(G_l^i + G_i^l) - \delta_{il} \tag{120}$$

For asymmetrical case, one has:

$$V^l \frac{\partial \omega_{il}}{\partial t} + V^l V^m \frac{\partial \omega_{lm}}{\partial X^i} = 0, \quad \omega_{il} = \frac{1}{2}(G_l^i - G_i^l) \tag{121}$$

They show that as each component of micro deformation tensor has different transportation speed. Very complicated time-space structures will be very common for turbulence. This fact has been studied with very long history. It seems that the turbulence is random as it is not deterministic. For vortex motion, the solution is deterministic. For symmetrical case, the solution is not completely deterministic. This is named as instability or multi-solution in many researches.

**6.3 Reynolds number for viscosity-sensitive turbulence**

For very small viscosity fluid, one has:

$$\left| \rho V^l \frac{\partial G_l^i}{\partial t} \right| >> \left| 2\mu F_l^m \frac{\partial F_m^l}{\partial X^i} \right| \tag{122}$$

The wave motion equation becomes equation Equ.(118). Therefore, the very high speed condition equation Equ.(117) is equivalent with the above condition.

As the macro deformation is finite (usually $F_j^i \approx \delta_j^i$), the classical Reynolds number Re is defined as:

$$\text{Re} = \frac{\rho V}{\mu} \quad (123)$$

Then, the turbulence wave production condition is: $\text{Re} \gg 1$. There are many researches on the intrinsic meaning of Reynolds number. Here, it is clear that at least two possible interpretations are acceptable.

On the other hand, for very large viscosity fluid, one has:

$$\left| \rho V^l \frac{\partial G_l^i}{\partial t} \right| \ll \left| 2\mu F_l^m \frac{\partial F_m^l}{\partial X^i} \right| \quad (124)$$

The wave motion equation becomes equation Equ.(114). Therefore, the very low speed condition equation Equ.(113) is equivalent with the above condition. This condition means that: $\text{Re} \ll 1$. The micro deformation is deterministic. No significant turbulence is produced.

Summering up above discussion, the conclusion is that: the Reynolds number $\text{Re} = \frac{\rho V}{\mu}$ indeed can be taken as the critical for turbulence wave production in "incompressible" fluid and/or viscosity sensitive fluid flow problems.

## 7. Wave equations of pressure-sensitive turbulence

Unlike the viscosity-sensitive turbulence, there exists pressure-sensitive turbulence. This topic is discussed in many researches but with very confusing concept. The main problem for them is the Reynolds number cannot be used directly. This topic is the focus of this section.

The pressure-sensitive turbulence is produced under the conditions $\lambda(F_l^l - 3) \ll p$ and $2\mu \ll p$. Its wave equation is:

$$\rho V^l \frac{\partial G_l^i}{\partial t} + p \cdot \frac{\partial F_l^l}{\partial X^i} + \rho V^l V^m \frac{\partial G_m^l}{\partial X^i} = 0 \quad (125)$$

Very low velocity and very high velocity cases will be discussed. Then, the invariant turbulence structure problem will be addressed in some details. The experiments will be examined when it is appropriate.

### 7.1 Low speed flow or high pressure flow

Low speed flow is defined by the condition:

$$\left| p \cdot \frac{\partial F_l^l}{\partial X^i} \right| \gg \left| \rho V^l V^m \frac{\partial G_m^l}{\partial X^i} \right| \quad (126)$$

Or:

$$\left| \rho V^l \frac{\partial G_l^i}{\partial t} \right| \gg \left| \rho V^l V^m \frac{\partial G_m^l}{\partial X^i} \right| \quad (127)$$

The equation is micro deformation generation or decaying equation:

$$\rho V^l \frac{\partial G_l^i}{\partial t} + p \cdot \frac{\partial F_l^l}{\partial X^i} \approx 0 \quad (128)$$

Omitting the product spatial variation of the velocity and the turbulence deformation rate, its approximated solution is:

$$\rho V^l G_l^i = -\int_0^t p \cdot \frac{\partial F_l^l}{\partial X^i} dt + \rho V^l G_l^i \Big|_{t=0} \quad (129)$$

It shows that: for expanding macro deformation, the micro deformation is decreased; for compressing macro deformation, the micro deformation is increased. Although the micro deformation tensor $G_j^i$ cannot be completely determined by the above equation, however, the instant velocity is locally deterministic. They are:

$$\widetilde{V}^i = -\frac{1}{\rho} \int_0^t p \cdot \frac{\partial F_l^l}{\partial X^i} dt + \widetilde{V}^i \Big|_{t=0} \quad (130)$$

It shows that the flow velocity is controlled by the static pressure, mass density, and macro deformation gradient. This solution is very valuable for industrial application. The micro deformation is very deterministic. There in no turbulence in classical sense.

### 7.2 High speed or low pressure flow

The high speed or low pressure flow is defined by the condition:

$$\left| p \cdot \frac{\partial F_l^l}{\partial X^i} \right| << \left| \rho V^l V^m \frac{\partial G_m^l}{\partial X^i} \right| \quad (131)$$

Or,

$$\left| p \cdot \frac{\partial F_l^l}{\partial X^i} \right| << \left| \rho V^l \frac{\partial G_l^i}{\partial t} \right| \quad (132)$$

The equation becomes turbulence wave equation:

$$V^l \frac{\partial G_l^i}{\partial t} + V^l V^m \frac{\partial G_m^l}{\partial X^i} = 0 \quad (133)$$

This kind of wave has anisotropy transportation speed. It has been studied in the previous section.

For the condition equations Equ.(132), the first kind of Reynolds number related with pressure can be defined as:

$$\text{Re}(p) = \frac{\rho V}{p} \quad (134)$$

Where, $V$ is the stream flow velocity. For $\text{Re}(p) >> 1$, the turbulence will be produced and transported. This phenomenon is common in low pressure high speed flow.

Based on above discussion, referring equation Equ.(132), for high speed flow, the first kind of Reynolds number related with pressure can be calculated as:

$$\text{Re}(p) = Mean \left( \frac{\rho V^l \cdot \frac{\partial G_l^i}{\partial t}}{p \frac{\partial F_k^k}{\partial X^i}} \right) \approx \frac{\rho V}{p} \quad (135)$$

For this formulation, the maximum of $\frac{\partial G_l^i}{\partial t}$ and $\frac{\partial F_k^k}{\partial X^i}$ are supposed as the same order infinitesimal. There ratio has the velocity dimension.

On the other hand, referring equation Equ.(126) and (131), for high speed flow, the second kind of Reynolds number related with pressure can be calculated as:

$$\mathrm{Re}(p,V) = Mean\left(\frac{\rho V^l V^m \frac{\partial G_m^l}{\partial X^i}}{p \cdot \frac{\partial F_l^l}{\partial X^i}}\right) \approx \frac{\rho V^2}{p} \tag{136}$$

For this formulation, the maximum of $\frac{\partial G_m^l}{\partial X^i}$ and $\frac{\partial F_k^k}{\partial X^i}$ are supposed as the same order infinitesimal.

Therefore, the Reynolds number related with pressure can be explained by two senses. The differences are expressed by a characteristic length parameter. Based on this research, it has the length dimension or says it is the scale ratio of micro deformation over macro deformation. This non-uniqueness causes many confuses in fluid dynamics researches. Here, its intrinsic meaning is exposed.

**7.3 Invariant phase**

The first kind of invariant phase can be defined as:

$$p \cdot \frac{\partial F_l^l}{\partial X^i} \approx -\rho V^l V^m \frac{\partial G_m^l}{\partial X^i} \tag{137}$$

This condition is the second kind of Reynolds number equal to 1. In this case, the micro deformation will be constant as the motion equation becomes:

$$\rho V^l \frac{\partial G_l^i}{\partial t} \approx 0 \tag{138}$$

This means that the turbulence structure is invariant about time. For spatial periodic macro deformation, the micro deformation will be spatial periodic also. In this case, the equation Equ.(137) gives out the wave number relationship between macro deformation and micro deformation.

The second kind of invariant phase can be defined as:

$$\rho V^l \frac{\partial G_l^i}{\partial t} + p \cdot \frac{\partial F_l^l}{\partial X^i} \approx 0 \tag{139}$$

This condition is the first kind of Reynolds number equal to minus 1. In this case, the micro deformation will be constant as the motion equation becomes:

$$V^l V^m \frac{\partial G_m^l}{\partial X^i} = 0 \tag{140}$$

This means that the turbulence structure is invariant about space. For spatial periodic macro deformation, the micro deformation will be time periodic. In this case, the equation Equ.(139) gives out the turbulence frequency and the macro deformation wave number relationship. Generally speaking, the higher the macro wave number is, the higher is the turbulence frequency.

So, for invariant phase, although the turbulence solution is not complete in theory, the wave number-frequency relationship or wave number-number relationship are deterministic in above sense.

*7.3.1 Time-invariant turbulence structure*

For one dimensional flow along $X^1$ direction, the time-invariant turbulence structure meets equation:

$$p \cdot \frac{\partial F_1^1}{\partial X^1} \approx -\rho V^1 V^1 \frac{\partial G_1^1}{\partial X^1} \tag{141}$$

It can be expressed by the second of Reynolds number as:

$$\frac{\partial G_1^1}{\partial X^1} = -\frac{p \cdot \frac{\partial F_1^1}{\partial X^1}}{\rho V^1 V^1} = -\frac{1}{\text{Re}(p,V)} \cdot \frac{\partial F_1^1}{\partial X^1} \tag{142}$$

It says that the macro deformation gradient spatial variation over the second kind of Reynolds number is the turbulence deformation gradient spatial variation.

If viewing the Reynolds number as a constant parameter as it is usually done in turbulence experiments data analysis, the form solution is:

$$G_1^1 = -\frac{1}{\text{Re}(p,V)} \cdot F_1^1 + g \tag{143}$$

Where, $g$ is an arbitral constant waiting to be determined by boundary conditions.

Based on this physical understanding, very low Reynolds number ($\text{Re}(p,V) \ll 1$) means that the very small macro deformation will be dramatically amplified into the large micro deformation. This should be named as micro spatial-instability.

For very high Reynolds number ($\text{Re}(p,V) \gg 1$), a very small micro deformation variation will produce significant macro deformation. This conclusion is supported by experiments. This should be named as macro spatial-instability.

### 7.3.2 Space-invariant turbulence structure

For one dimensional flow along $X^1$ direction, the space-invariant turbulence structure meets equation:

$$\rho V^1 \frac{\partial G_1^1}{\partial t} + p \cdot \frac{\partial F_1^1}{\partial X^1} \approx 0 \tag{144}$$

If viewing the Reynolds number as a constant parameter, it becomes equation:

$$\frac{\partial G_1^1}{\partial t} + \frac{1}{\text{Re}(p)} \cdot \frac{\partial F_1^1}{\partial X^1} \approx 0 \tag{145}$$

Its form solution is:

$$G_1^1 = -\frac{1}{\text{Re}(p)} \cdot \int \frac{\partial F_1^1}{\partial X^1} dt \tag{146}$$

For very low Reynolds number ($\text{Re}(p) \ll 1$), the macro deformation variation is dramatically amplified into micro deformation along with time elapsing. This phenomenon is named as time-instability in many researches.

The minus sign means that the micro expansion is produced by macro compressing and the micro compressing is produced by macro expansion. The former is mainly related with the bubble production processes, the later is mainly related with the typhoon or local condensation. Both phenomena cause many researches for their importance in industrial problem and safety engineering.

For very high Reynolds number ($\text{Re}(p) \gg 1$), the micro deformation will grow up steadily a lone the time increase. In fact, many well observed turbulences in experiments are produced by this way. Initially, the macro deformation has a very small variation (such as in the input port), as the Reynolds number is very high, there is no micro deformation be observed. After a long time, the micro deformation becomes significant and keeps growing up until develops into a complicated turbulence structure pattern.

### 7.3.3 Transition phase

Generally, the time-invariant condition and the space-invariant condition cannot be maintained at the same time.

When the one dimensional flow is very static without significant macro deformation and micro deformation, what will be happen if the following conditions are met at the same time:

$$\frac{\partial F_1^1}{\partial X^1} \approx \text{Re}(p,V) \cdot \frac{\partial G_1^1}{\partial X^1} \tag{147-1}$$

$$\text{Re}(p) \cdot \frac{\partial G_1^1}{\partial t} + \frac{\partial F_1^1}{\partial X^1} \approx 0 \tag{147-2}$$

In this case, one has:

$$\text{Re} \cdot (p) \cdot \frac{\partial G_1^1}{\partial t} + \text{Re}(p,V) \cdot \frac{\partial G_1^1}{\partial X^1} \approx 0 \tag{148}$$

Its form solution is:

$$G_1^1 = f(X^1 + \frac{\text{Re}(p,V)}{\text{Re}(p)} \cdot t) \tag{149}$$

Where, the $f(z)$ is an arbitral function waiting to be determined by the initial or boundary conditions. According to wave theory, the parameter $\text{Re}[2]/\text{Re} = \omega$ is the radial frequency for periodic function $f(z)$. The inward-traveling feature of micro deformation wave feature shows that all the boundary micro deformation and initial micro deformation will be stacked up at some where at some time. Therefore, the turbulence will seem to be suddenly grown-up at there. This is the main feature of turbulence production. There, a turbulence focus region is formed.

On the other hand, by the form solution equation Equ.(149), one has:

$$\frac{\partial G_1^1}{\partial X^1} = \frac{\text{Re}(p)}{\text{Re}(p,V)} \cdot \frac{\partial G_1^1}{\partial t} \tag{150}$$

So, one has:

$$\frac{\partial F_1^1}{\partial X^1} \approx \text{Re}(p) \cdot \frac{\partial G_1^1}{\partial t} \tag{151}$$

It says that the growth rate of micro deformation will produce the significant spatial variation of macro deformation. In form, the macro deformation can be expressed as:

$$F_1^1 \approx \text{Re}(p) \cdot \int \frac{\partial G_1^1}{\partial t} dX^1 \tag{152}$$

Once the turbulence focus region is formed, the macro deformation will be significantly changed. This change will cause further micro deformation, and new turbulence focus region will be formed. It means that: firstly, a regional turbulence is formed; secondly, the macro deformation is significantly changed; thirdly, the sharp boundary of focus region of micro turbulence causes the significant growth rate of micro turbulence, this causes the macro deformation forms a sharp boundary; fourth, the newly formed macro deformation boundary causes new turbulence region be formed. Then, the process is repeated again. Summering up above points, the turbulence grown-up region is very sensitive about the geometrical boundary condition, while the macro geometrical boundary is very sensitive about the turbulence grown-up rate. Therefore, the time pattern and space pattern seem to be in struggle.

This interacting process is named as the transition phase in many researches.

## 8. Turbulence wave equation for vortex flow

In many fluid motions, small vortex motion is very common. The vortex flows like a natural composition of fluids. By Stokes theory, the small vortex is expressed as:

$$\omega_{ij} = \frac{1}{2}(\frac{\partial V^i}{\partial X^j} - \frac{\partial V^j}{\partial X^i}) \tag{153}$$

For very small macro velocity gradient, its geometrical meaning is local rotation. If the macro velocity gradient is big enough, the local rotation cannot be exactly expressed by it. So, a more general definition is required for the exact vortex description. This work has been finished by Prof. Chen Zhi-da in 1980s [12]. The new results are named as Stokes-Chen additive decomposition. Here, the elated results will be used directly.

**8.1 Unit orthogonal local rotation.**

Based on Stokes-Chen additive decomposition of deformation tensor, the small shear deformation tensor can be decomposed as:

$$F^i_j = S^i_j + R^i_j \tag{154}$$

Where, $S^i_j$ is a symmetrical tensor which represents intrinsic stretching; $R^i_j$ is an unit orthogonal tensor which represents the local average rotation. As the $R^i_j = \delta^i_j$ case is studied in the previous section, here, the $S^i_j = 0$ case will be studied.

Geometrically, the local average rotation will not change the local gauge tensor, so it is a gauge invariant deformation. The unit orthogonal local rotation flow is defined as:

$$F^i_j = R^i_j, \quad R^l_i R^l_j = \delta_{ij} \tag{155}$$

Based on Chen theory, the related items can be expressed as:

$$S^i_j = \frac{1}{2}(\frac{\partial V^i}{\partial X^j} + \frac{\partial V^j}{\partial X^i}) - (1 - \cos\Theta)L^i_l L^l_j \tag{156-1}$$

$$R^i_j = \delta^i_j + L^i_j \sin\Theta + (1 - \cos\Theta)L^i_l L^l_j \tag{156-2}$$

Where, $L^i_j$ is the rotation direction tensor component; $\Theta$ is the rotation angular ($0 \leq \Theta < \pi/2$). They are defined as:

$$L^i_j = \frac{1}{\sin\Theta}\omega_{ij} = \frac{1}{\sin\Theta} \cdot \frac{1}{2}(\frac{\partial V^i}{\partial X^j} - \frac{\partial V^j}{\partial X^i}) \tag{156-3}$$

$$\Theta = \arcsin\sqrt{(\omega_{12})^2 + (\omega_{23})^2 + (\omega_{31})^2} \tag{156-4}$$

The unit local rotation tensor only has three independent components. Based on this formulation, the physical acceptability condition is:

$$\sqrt{(\omega_{12})^2 + (\omega_{23})^2 + (\omega_{31})^2} < 1 \tag{156-5}$$

For such a kind of flow, the kinetic energy is conservative. The classical symmetrical strain rate is produced by the local rotation and is not zero. Hence, the stress will not be zero.

By the condition $S^i_j = 0$, the classical symmetrical strain tensor is:

$$\varepsilon_{ij} = \frac{1}{2}(\frac{\partial V^i}{\partial X^j} + \frac{\partial V^j}{\partial X^i}) = (1 - \cos\Theta)L^i_l L^l_j \tag{157}$$

Hence, the classical symmetrical stress is:

$$\sigma^i_j = -p\delta^i_j - 2\lambda(1 - \cos\Theta)\delta^i_j + 2\mu(1 - \cos\Theta)L^i_l L^l_j \tag{158}$$

Under above definitions, for vortex flow, the general turbulence wave equation Equ.(109) is simplified as

$$\rho V^l \frac{\partial G_l^i}{\partial t} + 2 \cdot [p + 2\lambda(1-\cos\Theta) + 2\mu] \frac{\partial \cos\Theta}{\partial X^i} + \rho V^l V^m \cdot \frac{\partial G_m^l}{\partial X^i} = 0 \qquad (159)$$

This is the **general wave equation for the turbulence in vortex flow.**

Its very low speed approximation of the general turbulence wave equation Equ.(159) is:

$$\rho V^l \frac{\partial G_l^i}{\partial t} + 2 \cdot [p + 2\lambda(1-\cos\Theta) + 2\mu] \frac{\partial \cos\Theta}{\partial X^i} = 0 \qquad (160)$$

Hence, the low speed solution is:

$$\tilde{V}^i = \frac{1}{\rho} \int_0^t 2 \cdot [p + 2\lambda(1-\cos\Theta) + 2\mu] \frac{\partial \cos\Theta}{\partial X^i} dt + \tilde{V}^i \Big|_{t=0} \qquad (161)$$

It says that the local rotation angular gradient and fluid parameters determine the flow velocity variation. However, the turbulence deformation is not completely deterministic.

The very high speed approximation of the general turbulence wave equation Equ.(159) is:

$$\rho V^l \frac{\partial G_l^i}{\partial t} + \rho V^l V^m \frac{\partial G_m^l}{\partial X^i} = 0 \qquad (162)$$

It is identical with previous results for high speed approximation. The turbulence is purely an anisotropy wave.

The time-invariant phase is given out by the following equation:

$$2 \cdot [p + 2\lambda(1-\cos\Theta) + 2\mu] \frac{\partial \cos\Theta}{\partial X^i} + \rho V^l V^m \cdot \frac{\partial G_m^l}{\partial X^i} = 0 \qquad (163)$$

It says that the turbulence deformation tensor gradient is limited by the local rotation angular gradient and fluid parameters. In fact, it gives out the wave number relation between the macro deformation and micro deformation. Letting the macro deformation wave number vector is $k_i$ and the micro deformation wave number vector is $\tilde{k}_i$, then the micro spatial scale and macro spatial scale is related by the algebra equation:

$$\left| 2 \cdot [p + 2\lambda(1-\cos\Theta) + 2\mu] \cdot \cos\Theta \right| \cdot k_i = \left| \rho V^l V^m G_m^l \right| \cdot \tilde{k}_i \qquad (164)$$

This result shows that the spatial scale ratio between the vortex and turbulence is controlled by the velocity field, local rotation angular, and fluid parameters.

The spatial-invariant phase is given out by the following equation:

$$\rho V^l \frac{\partial G_l^i}{\partial t} + 2 \cdot [p + 2\lambda(1-\cos\Theta) + 2\mu] \frac{\partial \cos\Theta}{\partial X^i} = 0 \qquad (165)$$

It shows that the micro deformation period is determined by the macro deformation wave number. Letting the $\tilde{f}$ be the micro deformation frequency and the $k_i$ be the macro deformation wave number vector, then they are related by a simple algebra equation:

$$\left| \rho V^l G_l^i \right| \cdot \tilde{f} = \left| 2 \cdot [p + 2\lambda(1-\cos\Theta) + 2\mu] \cdot (\cos\Theta) \right| \cdot k_i \qquad (166)$$

There are many researches put their attention on establishing the turbulence frequency and wave number equation referring the macro deformation features and flow parameters. In this research, the relation is deterministic.

**8.2 Orthogonal local rotation with expansion**

In nature, when vortex motion is strong enough, bubbles will be produced. The material element volume is expanded in macro sense. Such a kind of bubbling vortex motion is expressed as the orthogonal local rotation with expansion, hereafter [25-26]. Theoretically, the orthogonal local rotation with expansion is defined as:

$$F^i_j = \tilde{S}^i_j + \frac{1}{\cos\theta} \cdot \tilde{R}^i_j \tag{167}$$

Where, $\tilde{S}^i_j$ is a symmetrical tensor; $\tilde{R}^i_j$ is an unit orthogonal rotation tensor with rotation angular $\theta$ ($0 \le \theta < \pi/2$).

The local rotation angular is defined as:

$$\left(\frac{1}{\cos\theta}\right)^2 = 1 + (\omega_{12})^2 + (\omega_{23})^2 + (\omega_{31})^2 \tag{168-1}$$

For such a kind of deformation, the intrinsic stretching is zero. The related equations are:

$$\tilde{S}^i_j = \frac{1}{2}\left(\frac{\partial V^i}{\partial X^j} + \frac{\partial V^j}{\partial X^i}\right) - \left(\frac{1}{\cos\theta} - 1\right)(\tilde{L}^i_k \tilde{L}^k_j + \delta^i_j) \tag{168-2}$$

$$\frac{1}{\cos\theta}\tilde{R}^i_j = \frac{1}{2}\left(\frac{\partial V^i}{\partial X^j} - \frac{\partial V^j}{\partial X^i}\right) + \left(\frac{1}{\cos\theta} - 1\right)(\tilde{L}^i_k \tilde{L}^k_j + \delta^i_j) \tag{168-3}$$

$$\tilde{R}^i_j = \delta^i_j + \sin\theta \cdot \tilde{L}^i_j + (1-\cos\theta)\tilde{L}^i_k \tilde{L}^k_j \tag{168-4}$$

$$\tilde{L}^i_j = \frac{\cos\theta}{2\sin\theta}\left(\frac{\partial V^i}{\partial X^j} - \frac{\partial V^j}{\partial X^i}\right) \tag{168-5}$$

It is clear that the deformation is completely determined by the vortex field.

The purely orthogonal local rotation with expansion is defined as: $\tilde{S}^i_j = 0$. In this case, the classical strain rate is:

$$\varepsilon_{ij} = \frac{1}{2}\left(\frac{\partial V^i}{\partial X^j} + \frac{\partial V^j}{\partial X^i}\right) = \left(\frac{1}{\cos\theta} - 1\right)(\tilde{L}^i_k \tilde{L}^k_j + \delta^i_j) \tag{169}$$

The classical stress tensor is:

$$\sigma^i_j = -p\delta^i_j + \lambda\left(\frac{1}{\cos\theta} - 1\right)\delta^i_j + 2\mu\left(\frac{1}{\cos\theta} - 1\right)(\tilde{L}^i_k \tilde{L}^k_j + \delta^i_j) \tag{170}$$

Under above definitions, for orthogonal local rotational flow with expansion, the general turbulence wave equation for simple fluid Equ.(109) is simplified as

$$\rho V^l \frac{\partial G^i_l}{\partial t} + \{[p - \lambda \cdot (\frac{1}{\cos\theta} - 1)] + 2\mu\} \cdot \frac{\partial(\frac{1}{\cos\theta})}{\partial X^i} - 2\mu \frac{\partial(\frac{1}{\cos\theta})^2}{\partial X^i} + \rho V^l V^m \frac{\partial G^l_m}{\partial X^i} = 0 \tag{171}$$

After an algebra operation, it can be rewritten as:

$$\rho V^l \frac{\partial G^i_l}{\partial t} + [p + \lambda + 2\mu - (\lambda + 4\mu) \cdot \frac{1}{\cos\theta}] \cdot \frac{\partial(\frac{1}{\cos\theta})}{\partial X^i} + \rho V^l V^m \frac{\partial G^l_m}{\partial X^i} = 0 \tag{172}$$

This is the **turbulence wave equation for orthogonal local rotation with expansion.**

Its very low speed approximation is:

$$\rho V^l \frac{\partial G^i_l}{\partial t} + [p + \lambda + 2\mu - (\lambda + 4\mu) \cdot \frac{1}{\cos\theta}] \cdot \frac{\partial(\frac{1}{\cos\theta})}{\partial X^i} = 0 \tag{173}$$

It shows that the turbulence motion is produced by volume expansion (that is the bubbling). It also shows that the micro deformation is controlled by the rotation angular of vortex and fluid parameters. For von Karman vortex street, if the wave number vector of macro vortex is $k_i$ and the turbulence frequency is $\tilde{f}$, then the turbulence frequency and vortex wave number are related by the following algebra equation:

$$\left| \rho V^l G_l^i \right| \cdot \tilde{f} = \left| [p + \lambda + 2\mu - (\lambda + 4\mu) \cdot \frac{1}{\cos\theta}] \cdot \frac{1}{\cos\theta} \right| \cdot k_i \quad (174)$$

It says that the turbulence frequency is determined by the macro deformation wave number and flow parameters.

There are many researches put their attention on establishing the turbulence frequency and wave number equation referring the macro deformation features and flow parameters. In this research, the relation is deterministic.

The time-invariant phase approximation is:

$$[p + \lambda + 2\mu - (\lambda + 4\mu) \cdot \frac{1}{\cos\theta}] \cdot \frac{\partial(\frac{1}{\cos\theta})}{\partial X^i} + \rho V^l V^m \frac{\partial G_m^l}{\partial X^i} = 0 \quad (175)$$

It shows that the micro deformation spatial feature is controlled by the macro deformation spatial variation. Letting the macro deformation wave number vector is $k_i$ and the micro deformation wave number vector is $\tilde{k}_i$, then the micro spatial scale and macro spatial scale is related by the algebra equation:

$$\left| \rho V^l V^m G_m^l \right| \cdot \tilde{k}_i = \left| [p + \lambda + 2\mu - (\lambda + 4\mu) \cdot \frac{1}{\cos\theta}] \frac{1}{\cos\theta} \right| \cdot k_i \quad (176)$$

It says that the turbulence wave number is determined by the macro deformation wave number and flow parameters.

There are many researches put their attention on establishing the turbulence frequency and wave number equation referring the macro deformation features and flow parameters. In this research, the relation is deterministic.

## 9. End-words

Although there are many researches and experimental results been published, it is difficult to collect direct detailed results which are comparable with the results in this research. However, as a summery for this paper, such a comparison will be helpful for understanding the value of the new theoretic formulation for turbulence problem. This is the main target of this section. As the published papers are overwhelming, only a very small portion is used here.

(1). The fluid flow is expressed by two deformation tensors: one is the average flow determined macro deformation, another is the turbulence determined micro deformation tensor. As the macro flow determines a curvature space, the turbulence flow happens on such a reference configuration. By this way, the fluid flow is exactly described by the two geometrical field tensors [12, 25-26].

(2). The deformation energy is viewed as a potential quantity as it depends only on the configuration, which is viewed as a kind of general coordinators [13]. Based on this point, the Lagrange quantity is defined in commoving dragging coordinator system. Based on Least Action principle, the Euler-Lagrange equations are obtained by coordinator variations [16-17], which can be expressed by the deformation tensors.

(3). The stress concept is introduced by extending the Green's definition [13]. By transforming the related differentiation about current coordinators into laboratory coordinator system, the spatial form of Euler-Lagrange equation expressed by deformation tensor and stress tensor are obtained.

(4). The traditional Navier-Stokes equation is obtained by simplifications or say the first order approximation. The effective ness of NS equation is evaluated. It shows that, the stress concept play the main role.

(5). In this research, the Bernoulli Equation is a natural result of general Euler-Lagrange motion equations. So, it is not fundamental law. This conclusion is partially supported as the Bernoulli Equation is not mentioned in many researches. In this research, it has been shown that the applicability condition of traditional fluid motion equation requires the Bernoulli Equation be effective. In physical sense, the Bernoulli Equation limits the stress concept for large deformation Equ.(88). If the Bernoulli Equation is abandon, the stress will become a modifiable concept. Modifying the stress concept is a very common phenomenon in many researches [27-29]. Unlike other theories, the Bernoulli Equation is implied by the general equation in this research. This is an important point, as the Bernoulli Equation is well supported by experiments [24, 30-31].

(6). After a brief general discussion, the motion equations of macro deformation without turbulence are established. This equation is firstly established in this research. It explains the main feature of turbulence phenomenon well. Taking the average fluid flow solution as the first approximation, the turbulence motion equations are obtained. In general sense, it is an inward-traveling non-linear wave equation. As a special case, turbulence wave solutions in one dimension flow are discussed. There are many visualization researches about the turbulence phenomenon in one dimension flow [32-47].

(7). For idea simple Newtonian fluid, the turbulence wave equation is expressed by the fluid parameters and macro fluid velocity field. The research shows that the turbulence wave equation can be classified into two kinds: one is viscosity-sensitive turbulence, another is pressure-sensitive turbulence.

(8). For viscosity-sensitive turbulence, the traditional Reynolds number definition is obtained. 1) For low speed flow, the turbulence is produced by the gradient of macro deformation tensor multiplied by viscosity parameters. The analytical turbulence solution form is obtained. 2) For very high speed flow, the turbulence wave equation is a linear inward-traveling wave equation. The velocity plays the role of phase velocity. This explains the standing wave feature of turbulence, which are more or less be avoided in many turbulence research papers. 3) The research confirms that for viscosity-sensitive turbulence, the classical Reynolds number plays an intrinsic role. Based on the related results, controlling the macro deformation by geometrical method (boundary condition), the turbulence feature can be controlled. There are some experimental researches can be used to support this kind of turbulence [32-47].

(9) For pressure-sensitive turbulence, the turbulence becomes a little more complicated. 1) For high pressure (low speed) flow, the turbulence is determined by the gradient of macro deformation tensor multiplied by pressure parameters. That is to say, the position of viscosity is replaced by pressure. The analytical turbulence solution form is obtained. 2) For low pressure (high speed), the turbulence wave equation is a linear inward-traveling wave equation. The velocity plays the role of phase velocity. This is similar with the viscosity-sensitive turbulence. 3) However, the Reynolds number should be redefined. So, generally speaking, the classical Reynolds number is not suitable for pressure-sensitive turbulence. 4) As a special feature, a standing phase may exist in pressure-sensitive turbulence. There are two typical patterns: one is time-invariant structure, another is spatial-invariant structure. Such a kind structure features are the main reasons against the stochastic point of turbulence. This is the first time to confirm the turbulence invariant-patterns with strict theoretical formulation. For pressure-sensitive turbulence, a little experimental researches are available [48-51].

(10) The vortex controlling turbulence phenomenon is formulated with explicit wave equations. The vortex is classified into two kinds: pure vortex and bulling vortex. The former in volume invariant, the later has volume expanding. The local rotation angular parameter is introduced to express the vortex feature. The obtained turbulence wave equation shows that: the gradient of local rotation angular of macro stream line is the force item for the turbulence wave equation. As the wave solution is boundary condition sensitive and highly depends on the initial condition, the research only gives up a simple discussion about the wave frequency and number relations. For industry application, as the stream line local rotation can be controlled by the turning of tube, these equations supply the basic wave equation for the turbulence control problem. So, they are valuable for industry application. In fact, there are many experimental researches about vortex and turbulence [52-61] or bubbling and turbulence [62-66].

(11) This research shows that the turbulence is a deterministic phenomenon. The main features which cause many confusing is its "inward-traveling" wave features. For such a kind of wave, focusing phenomenon makes the turbulence seems to be grown up automatically. Because, in many wave research fields, the inward-traveling wave is taken as not "physical admissible", the turbulence had not been viewed by this point. So, the turbulence wave equation obtained in this research opens a new field for further researches.

(12) As the spectrum of turbulence wave is the Fourier transformation of turbulence wave correlation, the related results in this research show that the statistic methods in turbulence research do have their physical bases. They are supported by this research. The experimental researches about the structure of turbulence are very rich, such as [67-73].

In one words, these results not only confirm the related turbulence researches, but also unifying the related research into the general Euler-Lagrange motion equations. Therefore, the inward-traveling turbulence wave equation lays a new base for turbulence motion research.

**Appendix 1 On Traditional Motion Equations**

Traditionally, the deformation energy is defined by the surface force acting on a closed deformed surface. For unit volume element, the net force item is expressed as:

$$T_j = \frac{\partial \sigma_j^i}{\partial X^i} \tag{A1-1}$$

For infinitesimal deformation, this force is produced by the acceleration force, hence, one has:

$$\frac{\partial \sigma_n^i}{\partial X^i} = \frac{\partial}{\partial t}(\rho V^n) + \rho V^j \frac{\partial V^n}{\partial X^j} \tag{A1-2}$$

This is the traditional motion equation. However, the net force concept for fluid is doubtful as the closed surface is an imaginary one. Although other ways can be used to get this equation more strictly, however, its physical meaning is the surface force action. This denies the deformation energy be stored in the internal motion of fluid. As the local rotation will not supply the correct net force, this equation only gives the motion relation for symmetrical stress components.

The motion equation (A1-2) can be obtained by the Lagrange mechanics. The Lagrange quantity is defined as:

$$L = \int_{t_0}^{t} [(\frac{1}{2}\rho V^i V^i) - U] \cdot dt \tag{A1-3}$$

Its variation is:

$$\delta L = \int_{t_0}^{t} [(\rho V^i \delta V^i) - \frac{\partial U}{\partial X^i} \delta X^i] \cdot dt \tag{A1-4}$$

As:

$$\delta V^i = \frac{\partial V^i}{\partial X^j} \delta X^j + \frac{\partial \delta X^i}{\partial t} \tag{A1-5}$$

The least action principle gives out:

$$\delta L = \int_{t_0}^{t} \{[(\rho V^i \frac{\partial V^i}{\partial X^j}) - \frac{\partial U}{\partial X^j}] - \frac{\partial (\rho V^j)}{\partial t}\} \delta X^j \cdot dt \tag{A1-6}$$

Hence, the motion equation is:

$$-\frac{\partial U}{\partial X^j} = \frac{\partial (\rho V^j)}{\partial t} - \rho V^i \frac{\partial V^i}{\partial X^j} \tag{A1-7}$$

Based on Green formulation, defined the stress sign properly, one has:

$$\frac{\partial U}{\partial X^j} = T_j = -\frac{\partial \sigma_j^i}{\partial X^i} \tag{A1-8}$$

The motion equation becomes:

$$\frac{\partial \sigma_j^i}{\partial X^i} = \frac{\partial (\rho V^j)}{\partial t} - \rho V^i \frac{\partial V^i}{\partial X^j} \tag{A1-9}$$

It is different from equation (A1-2) for the last item on the right side of the equation.

If the both equations be identical, the necessary condition is:

$$\rho V^j \frac{\partial V^n}{\partial X^j} = -\rho V^j \frac{\partial V^j}{\partial X^n} \tag{A1-10}$$

That is:

$$V^j (\frac{\partial V^n}{\partial X^j} + \frac{\partial V^j}{\partial X^n}) = 0 \tag{A1-11}$$

In traditional formulation, as the strain rate is defined as:

$$\varepsilon_{jn} = \frac{1}{2}(\frac{\partial V^n}{\partial X^j} + \frac{\partial V^j}{\partial X^n}) \tag{A1-12}$$

Geometrically, for purely anti-symmetrical deformation flow, the equation (A1-11) can be met. For symmetrical deformation, it means that the kinetic energy density variation about spatial position can be omitted. Both conditions are too artificial.

Based on above discussion, the Navier-Stokes fluid motion equation does not meet the Least Action principle in general case. So, the Navier-Stokes fluid motion equation is only an approximation for pure anti-symmetrical deformation flow or when the kinetic energy density variation about spatial position can be omitted. This may be the essential reason for the difficulty on solving turbulence problem by Navier-Stokes equation.